\documentclass[11pt]{article}
\usepackage[letterpaper,left=0.6in,right=0.6in,top=1in,bottom=1in]{geometry}

\usepackage{graphicx}
\usepackage{physics}
\usepackage{amsmath}
\usepackage{amssymb}
\usepackage{amsthm}
\usepackage{enumitem}
\usepackage{framed}

\usepackage{caption}
\usepackage{subfig}

\usepackage{adjustbox}
\usepackage{multirow}
\usepackage{multicol}
\usepackage{listings}
\usepackage{tikz}
\usepackage{xcolor}
\usepackage{bm}
\usepackage{mathtools}
\usepackage{algorithm}
\usepackage{algpseudocode,eqparbox}
\usepackage[T1]{fontenc}
\usepackage{wrapfig}
\usetikzlibrary{arrows,decorations.pathmorphing,decorations.footprints,decorations.pathreplacing,fadings,calc,trees,mindmap,shadows,decorations.text,patterns,positioning,shapes,matrix,fit}
\usetikzlibrary{shapes.misc}

\mathtoolsset{showonlyrefs}
\allowdisplaybreaks

\usepackage[colorlinks,citecolor=red,urlcolor=blue,bookmarks=false,hypertexnames=true]{hyperref}
\usepackage{anyfontsize}
\usepackage{letltxmacro}
\LetLtxMacro{\originaleqref}{\eqref}
\renewcommand{\eqref}{Eq.~\originaleqref}

\definecolor{codegreen}{rgb}{0,0.6,0}
\definecolor{codegray}{rgb}{0.5,0.5,0.5}
\definecolor{codepurple}{rgb}{0.58,0,0.82}
\definecolor{backcolour}{rgb}{0.95,0.95,0.92}

\usepackage{color}
\definecolor{bitcolor}{rgb}{0.99,0.93,0.0}
\definecolor{checkcolor}{rgb}{0.52941,0.80784,1}
\definecolor{channelcolor}{rgb}{0.67,0.88,0.69}
\definecolor{infocolor}{rgb}{0.82,0.62,0.91}
\definecolor{cqcolor}{rgb}{0.99,0.56,0.67}
\lstdefinestyle{mystyle}{
    backgroundcolor=\color{backcolour},   
    commentstyle=\color{codegreen},
    keywordstyle=\color{magenta},
    numberstyle=\tiny\color{codegray},
    stringstyle=\color{codepurple},
    basicstyle=\ttfamily\footnotesize,
    breakatwhitespace=false,         
    breaklines=true,                 
    captionpos=b,                    
    keepspaces=true,                 
    numbers=left,                    
    numbersep=5pt,                  
    showspaces=false,                
    showstringspaces=false,
    showtabs=false,                  
    tabsize=2
}

\usepackage{stmaryrd}

\usepackage[style=ieee,citestyle=numeric-comp,backend=biber,doi=true,url=false,eprint=true,minbibnames=2]{biblatex}

\AtEveryBibitem{\iffieldundef{doi}
      {}
      {\clearfield{eprint}\clearfield{eprinttype}\clearfield{eprintclass}}}

\usepackage{ifthen}
\usepackage{authblk}

\newif\ifitw
\itwtrue

\newif\ifextra
\extrafalse

\usepackage{mathtools}

\usetikzlibrary{shapes.geometric, arrows}
\tikzstyle{startstop} = [rectangle,  minimum width=3cm, minimum height=1cm,text centered,text width=10cm, draw=black ,fill=gray!20]
\tikzstyle{process} = [rectangle, minimum width=3cm, minimum height=1cm, text centered,text width=10cm, draw=black,fill=orange!20]
\tikzstyle{arrow} = [thick,->,>=stealth]
\tikzstyle{state}=[shape=circle,draw=blue!50,fill=blue!20]
\tikzstyle{observation}=[shape=rectangle,draw=orange!50,fill=orange!20]
\tikzstyle{lightedge}=[<-,dotted]
\tikzstyle{mainstate}=[state,thick]
\tikzstyle{mainedge}=[<-,thick]
\usepackage{color}
\definecolor{bitcolor}{rgb}{1,0.84314,0}
\definecolor{checkcolor}{rgb}{0.52941,0.80784,1}
\usepackage{pgfplots}
\pgfplotsset{compat=1.18}
\renewcommand{\epsilon}{\varepsilon}

\newtheorem{theorem}{Theorem}

\newtheorem{rem}[theorem]{Remark}
\newtheorem{lem}[theorem]{Lemma}
\newtheorem{defn}[theorem]{Definition}

\newcommand{\cD}{\mathcal{D}}
\newcommand{\cE}{\mathcal{E}}

\newcommand{\cG}{\mathcal{G}}

\newcommand{\cN}{\mathcal{N}}

\newcommand{\cQ}{\mathcal{Q}}

\newcommand{\cS}{\mathcal{S}}

\newcommand{\cV}{\mathcal{V}}

\newcommand{\ba}{\bm{a}}

\newcommand{\be}{\bm{e}}

\newcommand{\bp}{\bm{p}}

\newcommand{\br}{\bm{r}}

\newcommand{\bv}{\bm{v}}

\newcommand{\sym}[1]{S_N}

\renewcommand{\Pr}{\ensuremath{\mathbb{P}}}

\global\long\def\Pr{\mathbb{P}}

\newcommand{\cnop}{\boxast}

\newcommand{\mone}{\bm{1}}
\newcommand{\mzero}{\bm{0}}

\newcommand{\cx}{C_{X}}
\newcommand{\cz}{C_{Z}}
\newcommand{\bbF}{\mathbb{F}}
\newcommand{\symp}{\mathrm{s}}

\definecolor{resolved}{rgb}{0,0.5,0.45}
\newenvironment{resolved}{\begingroup\color{resolved}}{\endgroup}
\definecolor{proposed}{rgb}{0.85,0.35,0}

\providecommand{\symp}{\mathrm{symp}}

\makeatletter
\long\def\thanksii#1{\stepcounter{footnote}\protected@xdef\@thanks{\@thanks
    \protect\thanksiifootnotetext{\thefootnote}{#1}}}
\long\def\thanksiifootnotetext#1#2{\begingroup
    \def\@thefnmark{#1}\@footnotetext{#2}\endgroup}
\makeatother

\begin{document}
\title{Area Theorems for Quantum Error Correction\thanksii{This research was presented in part at the 2026 IEEE Information Theory Workshop (ITW) in Tempe, Arizona~\cite{Mandal-itw26}.}}
\author[1,2]{Avijit Mandal\thanks{ avijit.mandal@duke.edu}}
\author[1,2]{Mert G\"okduman}
\author[1,2,3]{Henry D. Pfister}
\affil[1]{Duke Quantum Center, Duke University}
\affil[2]{Department of Electrical and Computer Engineering, Duke University}
\affil[3]{Department of Mathematics, Duke University}
\date{}

\maketitle

\begin{abstract}
The extrinsic information transfer (EXIT) and generalized EXIT (GEXIT) area theorems provide fundamental conservation laws that connect the rate of classical codes to the performance of maximum-a-posteriori decoding on binary memoryless channels.
  In this work, we generalize these area theorems to quantum error correction.
  We first develop a coset-entropy formalism for nested linear codes, where the message specifies a coset and encoder randomization selects a codeword within that coset.
  The area under the resulting coset (G)EXIT function equals the normalized dimension of the quotient.
  For stabilizer codes, the measured syndrome restricts the Pauli error but does not always determine its logical effect, and the remaining uncertainty is an entropy over logical correction classes.
  We show that this entropy equals a classical coset entropy associated with the stabilizer group and the Pauli operators that commute with it.
  The resulting quantum GEXIT function has area equal to the number of encoded qubits per channel use for any stabilizer code, and this applies to the depolarizing channel through its joint four-valued Pauli-error distribution.
  For CSS codes on the quantum erasure channel and under independent bit- and phase-flip noise, the logical correction problem separates into two binary coset-decoding problems.
  On the quantum erasure channel, the duality between the stabilizer group and the Pauli operators that commute with it gives a folding identity at complementary erasure probabilities.
Using this and the area theorem, we derive an upper bound on the rate of code families with vanishing logical error probability that matches the capacity of the quantum erasure channel.
For the depolarizing channel, we also derive a rate-dependent converse, using a connection between correction-class entropy and coherent information to establish an inequality involving the complementary channel.

\end{abstract}
\section{Introduction}

Extrinsic information transfer (EXIT) curves were first introduced by ten Brink as a practical method to understand the flow of soft information in iterative decoders.
In his formulation, the curve plots the extrinsic mutual information between a code bit and the updated soft-estimate from a component decoder as a function of the mutual information between a code bit and the prior soft information supplied to the component decoder~\cite{tenBrink-com01}.
EXIT charts have since become a standard tool for analyzing and
designing turbo codes, serially concatenated codes, LDPC codes, and more
general code ensembles~\cite{tenBrink-com01,Ashikhmin-it04,Richardson-it01,Liva-css06,ElHajjar-comst14}.

A central property of EXIT functions is the area theorem, which follows from their interpretation as entropy derivatives.
For the binary erasure channel (BEC), the EXIT function is the normalized derivative of the conditional codeword entropy with respect to the erasure probability.
Its integral is fixed by the entropy difference between the noiseless and fully erased channels and therefore equals the code rate~\cite{Ashikhmin-it04}.
Related derivative identities connect the derivative of mutual information with respect to the signal-to-noise ratio to the minimum mean-square error~\cite{Guo-it05}, and GEXIT functions extend the entropy-derivative formulation and its area theorem to more general channel families~\cite{Measson-it09,Kudekar-it13,Kumar-it14}.
These area theorems relate the code rate to maximum-a-posteriori (MAP) decoding thresholds and are particularly useful for comparing them with belief-propagation (BP) thresholds~\cite{Kudekar-it11,Yedla-it14,Kudekar-it15,Kudekar-it17}.
EXIT charts have also been used to design iterative quantum turbo decoders and to compare their thresholds with the quantum hashing bound, but that work does not develop an area theorem for quantum error correction~\cite{Babar-vt15}.

The goal of this paper is to develop analogous area theorems for quantum stabilizer codes and to use them to derive rate constraints for the quantum erasure channel, independent bit- and phase-flip noise, and the depolarizing channel.
The key quantity is the correction-class entropy of a stabilizer code, which measures the uncertainty that remains in the logical correction after observing the syndrome and, for the erasure channel, the locations of the erasures.

We first formulate a classical coset-entropy area theorem for nested binary linear codes \(C_2\subseteq C_1\).
A message specifies a coset in \(C_1/C_2\), while encoder randomization selects a codeword from that coset.
After the noisy observation, the conditional entropy of the coset index measures the remaining uncertainty about the message, and its derivative with respect to the channel parameter defines a coset (G)EXIT function.
The area under this function equals the dimension of the quotient normalized by the blocklength.
This extends the usual codeword area theorem from decoding codewords to decoding cosets, and it is the natural classical precursor of stabilizer decoding, where logical corrections are determined only up to an element of the stabilizer group.

To extend this construction to quantum error correction, we represent Pauli errors up to phase by vectors over the additive alphabet $\bbF_2^2$, so that a Pauli channel acts by addition.
Let $\cS$ be the stabilizer group and let $\cS^\perp\supseteq\cS$ be the set of Pauli operators that commute with every element of $\cS$.
Errors with the same syndrome differ by an element of $\cS^\perp$, while errors that differ by an element of $\cS$ have the same effect on the encoded state.
Once the syndrome is known, the decoder needs to still determine the error modulo $\cS$, and the distinct logical corrections correspond to the cosets of $\cS$ in $\cS^\perp$.
The conditional entropy of this logical correction class is the central quantity studied in this paper.
We prove that it equals the coset entropy for the nested pair $\cS\subseteq\cS^\perp$ when the Pauli-error vector is added to a uniform element of $\cS^\perp$.
Thus, syndrome decoding for a stabilizer code is exactly a classical coset-decoding problem, and the information-theoretic analysis can be carried out using the posterior distribution of a classical coset observation.
Tracking this entropy as the normalized channel entropy increases defines a quantum GEXIT function whose area equals the number of encoded qubits per channel use.

For CSS codes on the quantum erasure channel, the bit- and phase-error components give two binary erasure coset-decoding problems with the same erasure set.
The same separation holds under independent bit- and phase-flip noise.
In both cases, the correction-class entropy decomposes into separate $X$-type and $Z$-type terms, and the quantum (G)EXIT function is the average of two classical coset (G)EXIT functions.

For the depolarizing channel, the bit- and phase-error components are correlated, so we use their joint four-valued distribution.
The channel applies each nonidentity Pauli error with probability $p/3$, and the entropy of this distribution increases from zero to two bits as $p$ increases from zero to $3/4$.
We parameterize the channel by half of this entropy, so that the parameter ranges from zero to one.
Since a depolarizing channel with a larger parameter can be obtained by following one with a smaller parameter by an independent depolarizing channel, the correction-class entropy is nondecreasing in the parameter.
The same degradation argument shows that changing the channel on one coordinate changes this entropy by at most two bits.
After normalization, the quantum GEXIT function therefore lies between zero and one, and its area equals the quantum code rate.

The area theorem fixes the total increase of the correction-class entropy between a noiseless and a completely noisy channel, but not the noise levels at which this increase occurs, and these levels determine up to which noise level the logical correction class can be recovered reliably.
For the quantum erasure channel, the relation between $\cS$ and $\cS^\perp$ gives a folding identity between the correction-class entropies at erasure probabilities $\epsilon$ and $1-\epsilon$.
Combined with the area theorem, this identity shows that any stabilizer code family whose optimal syndrome-decoding error vanishes at erasure probability $\epsilon$ with $0\leq \epsilon\leq 1/2$ has asymptotic rate $R\le 1-2\epsilon$, which equals the capacity of the quantum erasure channel~\cite{Bennett-physrevlett97}.
This argument is closely related to the rank-based analysis of stabilizer codes on the quantum erasure channel by Delfosse and Z\'emor, which controls the same correction-class dimensions without differentiating an entropy function~\cite{Delfosse-qic13}.

The final converse concerns the depolarizing channel.
We express the correction-class entropy in terms of the coherent information between the encoded state and the channel output, whose complementary output contains the information that is unavailable to the receiver.
By showing that part of this complementary output is itself the output of another depolarizing channel, we obtain a folded inequality between the correction-class entropies at two depolarizing parameters.
Applying Fano's inequality, this gives a rate-dependent converse for optimal syndrome decoding.
The resulting bound reaches zero at $p=1/4$, where the receiver output can be generated from the complementary output and the channel becomes antidegradable~\cite{Cubitt-jmp08}.
These area constraints may also help compare practical quantum decoding algorithms with optimal decoding~\cite{Pech-isit26,Freire-arxiv26,Yao-isit24,beni2025tesseract,leverrier2026approximating}.

The remainder of this paper is organized as follows.
Section~\ref{sec:background} introduces coset entropy, and Section~\ref{sec:coset} develops EXIT and GEXIT area theorems for coset coding on the BEC and the binary symmetric channel (BSC).
Section~\ref{sec:stabilizer} shows that syndrome decoding is coset decoding and proves the stabilizer area theorems for the quantum erasure and depolarizing channels, including the CSS erasure specialization.
Section~\ref{sec:converse} derives the converses, and Section~\ref{sec:numerical} presents numerical results.

\section{Coset Entropy}\label{sec:background}

Let $C\subseteq\bbF_2^n$ be a binary linear code or $C\subseteq(\bbF_2^2)^n$ be an additive code, and define the subcode supported on the index set $E\subseteq[n]$ by
\begin{align}\label{eq:C(E)}
    C(E)\coloneqq \{ c\in C \mid c_j =0 \text{ for all } j\notin E \}.
\end{align}

Consider binary linear codes $C_2, C_1$ satisfying 
\begin{align*}
    C_2\subseteq C_{1}\subseteq \mathbb{F}_2^{n}.
\end{align*}
Let $T$ be a complete set of coset representatives satisfying
\begin{align}\label{eq:coset decomposition}
    C_1=\bigsqcup_{m\in T}m+C_2.
\end{align}

The quotient $C_1/C_2$ partitions $C_1$ into the disjoint cosets $m+C_2$ indexed by $m\in T$.
This leads to a coset-coding scheme in which the two components play distinct roles.
The random variable $M$ selects the coset and represents the message, while an independent random variable $V$, uniform on $C_2$, selects a codeword within that coset.
The transmitted vector is $U=M+V$, and when $M$ and $V$ are uniform, $U$ is uniform on $C_1$.

After observing the channel output $Y$, the quantity $H(M\mid Y)$ measures the uncertainty about the transmitted coset.
It is zero when the observation determines the coset and equals $\log_2|T|$ when $M$ is uniform and the observation gives no information about it.
This differs from the codeword entropy $H(U\mid Y)$ because uncertainty among codewords in the same coset does not contribute to message uncertainty.
This distinction is central to both coset coding and stabilizer decoding.

\begin{lem}\label{lem:entropy-difference}
For a uniform vector $U\in C_1$, let $U=M+V$ be the unique decomposition of $U$ into $M\in T$ and $V \in C_2$.
Let $Y$ and $Y'$ be the outputs when $U$ and $V$, respectively, are transmitted through the same memoryless channel whose $j$th use is a binary memoryless symmetric (BMS) channel $W_j$.
Then, $M$ and $V$ are independent and uniform, and
    \begin{align*}
        H(U\mid Y)= H(M\mid Y)+H(V\mid Y').
    \end{align*}
\end{lem}

\begin{proof}
The map $(m,v)\mapsto m+v$ is a bijection from $T\times C_2$ to $C_1$.
Since $U$ is uniform, each pair has probability $1/|C_1|=1/(|T||C_2|)$.
Thus, $M$ and $V$ are independent and uniform. The entropy chain rule gives
\begin{align*}
H(U\mid Y)
&=H(M,V\mid Y)\\
&=H(M\mid Y)+H(V\mid M,Y).
\end{align*}
For a fixed $m\in T$ and each coordinate $j$ with $m_j=1$, the symmetry of $W_j$ provides an invertible output map that exchanges the output distributions for inputs $0$ and $1$.
Apply these maps to $Y$ and denote the resulting vector by $Y^{(m)}$.
Conditional on $M=m$, the pair $(V,Y^{(m)})$ has the same distribution as $(V,Y')$.
Invertibility and the independence of $M$ and $V$ therefore give
\begin{align*}
H(V\mid M=m,Y)
=H(V\mid M=m,Y^{(m)})
=H(V\mid Y').
\end{align*}
Averaging over $m$ proves the identity.
In particular, it applies after conditioning on a fixed BEC erasure set.
\end{proof}

\section{(G)EXIT Area Theorems for Coset Coding}\label{sec:coset}

\subsection{Related Work}

Area theorems for coset coding are closely connected to wiretap channels~\cite{Wyner-it75} and information-theoretic secrecy~\cite{Bloch-jsait21}.
In erasure and type-II wiretap problems, a confidential message identifies a coset, encoder randomization selects a codeword within that coset, and the eavesdropper's remaining uncertainty is measured by the equivocation $H(M\mid Y)$~\cite{Ozarow-bell84}.
LDPC constructions for wiretap channels were analyzed through equivocation and related coset-entropy quantities~\cite{Thangaraj-it07}, with subsequent work relating strong secrecy to decoding properties of dual LDPC codes~\cite{Subramanian-ifs11} and analyzing two-edge-type ensembles through equivocation and density evolution~\cite{Rathi-it13}.

EXIT charts have also been used to predict thresholds of iterative attacks in tandem coding and cryptographic wiretap models~\cite{Harrison-itw09,Harrison-isit09}.
Later finite-blocklength work studied equivocation and nested-code performance~\cite{Pfister-icc17,Harrison-aller18,Harrison-isit19,Harrison-comlett20}, while other work optimized nested linear secrecy codes using finite-blocklength criteria~\cite{Shoushtari-entropy23,Hunn-arxiv24}.

In this paper, the conditional entropy $H(M\mid Y)$ plays a different role.
Its derivative defines a coset (G)EXIT function whose integral is determined by the dimension of the quotient, independently of any particular decoder.
This provides the classical formulation used in Section~\ref{sec:stabilizer}, where a logical correction class of a stabilizer code takes the role of the coset message.

\subsection{EXIT Functions on the BEC}

For a code $C\subseteq \mathbb{F}_2^n$, let $H^{C}\in \mathbb{F}_{2}^{r\times n}$ denote a parity-check matrix satisfying $C=\ker H^{C}$. Throughout we assume parity-check matrices have full row rank. For a subset $\cE\subseteq[n]$, let $H^C_{\cE}$ denote the submatrix of $H^C$ formed by the columns indexed by $\cE$.
Consider a uniformly chosen codeword $U\in C$ which is transmitted through a BEC $W$ and let $Y$ be the channel output.
Let $E\subseteq[n]$ be the random erasure set whose elements are chosen independently with $\Pr(j\in E)=\epsilon_j$. For a fixed realization $E=\cE$, the output is
\begin{align*}
    Y_j=\begin{cases}
        U_j, \qquad \text{if $j\notin \cE$}\\
        ?, \qquad \text{otherwise}
    \end{cases}
\end{align*}
In this case, the parity-check equation becomes 
\begin{align*}
H^{C}_{\cE}U_{\cE}+H^{C}_{\overline{\cE}}U_{\overline{\cE}}=\mzero,
\end{align*}
where $\overline{\cE}\coloneqq[n]\setminus \cE$ is the relative complement of $\cE$.
Conditional on \(E=\cE\) and \(Y\), the codeword $U$ must lie in an affine subspace defined by the above equation.
This subspace has dimension \(|\cE|-\rank H^C_{\cE}\) and must contain the true codeword.
Thus, we get
\begin{align}\label{eq:bec-entropy}
    H(U\mid E=\cE,Y) & = |\cE|-\rank H^{C}_{\cE}
    =\dim C(\cE),
\end{align}
where $\cE$ is some fixed erasure index set and the second equality follows from the fact that $C=\ker H^C$. 

Consider the erasure probability vector $\bm{\epsilon}=[\epsilon_1,\dots,\epsilon_n]\in [0,1]^n$, where $\epsilon_j$ denotes the erasure probability for the $j$th channel input.
Then, the set of erasures $E$ satisfies \[\Pr(E=\cE) = \prod_{j\in \cE}\epsilon_j\prod_{j\notin \cE}(1-\epsilon_j). \]
For vector $\bm{\epsilon}\in [0,1]^n$, define the conditional entropy 
\begin{align*}
    A_{C}(\bm{\epsilon})\coloneqq H(U\mid E,Y) = \mathbb{E}[\dim C(E)]
\end{align*}
and the shorthand $A_{C}(\epsilon) \coloneqq A_{C}(\epsilon,\dots,\epsilon)$.

\begin{defn}[{\cite{Ashikhmin-it04}}] \label{def:exit_bec}
Let \(U\) be uniformly distributed over a binary linear code \(C\subseteq \bbF_2^n\) and $Y$ be the output of a BEC with input $U$ and erasure probability \(\epsilon\).
Then, the normalized EXIT function is defined to be
\begin{align*}
    h_{C}(\epsilon) \coloneqq \frac{1}{n}\sum_{j=1}^{n} H(U_j\mid E_{\sim j},Y_{\sim j}).
\end{align*}
\end{defn}

\begin{theorem}[{\cite{Ashikhmin-it04}}] \label{thm:exit_area}
Under the setup in Definition~\ref{def:exit_bec}, we have
\[
  h_{C}(\epsilon) = \frac{1}{n} \frac{d}{d\epsilon} A_C(\epsilon). 
\]
More generally, for per-coordinate erasure probabilities $\bm\epsilon\in[0,1]^n$, we have $\frac{\partial A_C(\bm\epsilon)}{\partial\epsilon_j}=H(U_j\mid E_{\sim j},Y_{\sim j})$ for each $j$.
It follows that
\[
        \int_0^1 h_C(\epsilon)\,d\epsilon
        =
        \frac{A_C(1)-A_C(0)}{n}
        =
        \frac{\dim C}{n}.
\]
\end{theorem}

\begin{proof}
This was established in~\cite{Ashikhmin-it04}.
For completeness, we include a proof in Appendix~\ref{app:known-proofs}.
\end{proof}

\subsection{Coset Coding on the BEC}

Consider nested codes $C_2\subseteq C_1\subseteq\mathbb F_2^n$, and let $T$ be a set of coset representatives satisfying \eqref{eq:coset decomposition}. Let $H^{C_1}\in\mathbb F_2^{r_1\times n}$ and $H^{C_2}\in\mathbb F_2^{r_2\times n}$ be parity-check matrices for $C_1$ and $C_2$, respectively. Let $U$ and $V$ be uniform on $C_1$ and $C_2$, respectively, and let $Y$ and $Y'$ be their BEC outputs under the same erasure set $E$. Then, Lemma~\ref{lem:entropy-difference} gives
\begin{align}\label{eq:bec-entropy-difference}
    H(M\mid E=\cE,Y)
    & =H(U\mid E=\cE,Y)-H(V\mid E=\cE,Y'),
\end{align}
where $M\in T$ is defined by $U=M+V$. Averaging gives
\begin{align*}
    H(M\mid E,Y)&=H(U\mid E,Y)-H(V\mid E,Y') \\
     & =\mathbb{E}[\rank H^{C_2}_{E}-\rank H^{C_1}_E].
\end{align*}
For a random erasure set $E$, drawn according to $\bm{\epsilon}$, define 
\begin{align*}
    A_{C_1/C_2}(\bm{\epsilon})\coloneqq H(M\mid E,Y),
\end{align*}
and the shorthand $A_{C_1/C_2}(\epsilon)\coloneqq A_{C_1/C_2}(\epsilon,\ldots,\epsilon)$.
Then, from \eqref{eq:bec-entropy-difference}, we get 
\begin{align*}
    A_{C_1/C_2}(\bm{\epsilon}) & = H(U\mid E,Y)-H(V\mid E,Y')\\
    & = A_{C_1}(\bm{\epsilon})- A_{C_2}(\bm{\epsilon}).
\end{align*}

\begin{defn} \label{def:coset_exit_bec}
For the coset coding problem on the BEC with $C_2\subseteq C_{1}$, the normalized coset EXIT function is defined by
\begin{align*}
    h_{C_1/C_2}(\epsilon)
    & \coloneqq \frac{1}{n}\sum_{j=1}^{n}\left(  H(U_j\mid E_{\sim j},Y_{\sim j})- H(V_j\mid E_{\sim j},Y'_{\sim j})\right).
\end{align*}
\end{defn}

\begin{theorem} \label{thm:coset_exit_area}
Under the setup in Definition~\ref{def:coset_exit_bec}, we have
\[
  h_{C_1/C_2}(\epsilon) = \frac{1}{n} \frac{d}{d\epsilon} A_{C_1/C_2} (\epsilon). 
\]
In addition, $A_{C_1/C_2}(\epsilon)$ is nondecreasing and Lipschitz with $0\leq h_{C_1/C_2}(\epsilon)\leq 1$.
It follows that
\[
        \int_0^1 h_{C_1/C_2} (\epsilon)\,d\epsilon
        =
        \frac{\dim C_1 - \dim C_2}{n}.
\]
\end{theorem}

\begin{proof}
Lemma~\ref{lem:entropy-difference} gives
$A_{C_1/C_2}(\bm\epsilon)=A_{C_1}(\bm\epsilon)-A_{C_2}(\bm\epsilon)$.
Differentiating on the diagonal and applying Theorem~\ref{thm:exit_area} proves
\begin{align*}
\frac{1}{n}\frac{d}{d\epsilon}A_{C_1/C_2}(\epsilon)
=h_{C_1}(\epsilon)-h_{C_2}(\epsilon)
=h_{C_1/C_2}(\epsilon).
\end{align*}
To verify the slope bounds directly, fix $j$ and repeat the proof of Theorem~\ref{thm:exit_area} with $M$ in place of $U$.
This gives
\begin{align}\label{eq:final1-bec-coset-slope}
\frac{\partial A_{C_1/C_2}(\bm\epsilon)}{\partial\epsilon_j}
=I(M;U_j\mid E_{\sim j},Y_{\sim j}).
\end{align}
Since $U_j$ is binary, the right side lies in $[0,1]$.
Hence, the coset entropy is nondecreasing and $1$-Lipschitz in each coordinate.
On the diagonal it is $n$-Lipschitz, and its normalized derivative lies in $[0,1]$.
At the endpoints,
\begin{align*}
A_{C_1/C_2}(0)=0,\qquad
A_{C_1/C_2}(1)=H(M)=\log_2|T|=\dim C_1-\dim C_2.
\end{align*}
The coset entropy is a difference of polynomials, so integrating its derivative proves the area identity.

\end{proof}

\begin{resolved}
\end{resolved}

The coset entropy of $C_1/C_2$ at erasure probability $\epsilon$ and that of the dual nested pair $C_2^\perp/C_1^\perp$ at $1-\epsilon$ sum to $\dim C_1-\dim C_2$.
The following lemma establishes this relation using the duality between code shortening and puncturing.

\begin{lem}\label{lem:bec-coset-duality}
Let $C_2\subseteq C_1\subseteq\bbF_2^{n}$ and let $\mone-\bm{\epsilon}$ denote the complementary erasure vector.
Then, we have
\begin{align}\label{eq:bec-folding}
    A_{C_1/C_2}(\bm{\epsilon})+A_{C_{2}^{\perp}/C_{1}^{\perp}}(\mone-\bm{\epsilon})
    =\dim C_1-\dim C_2 .
\end{align}
Consequently, the coset EXIT functions satisfy
\begin{align*}
    h_{C_1/C_2}(\epsilon)=h_{C_{2}^{\perp}/C_{1}^{\perp}}(1-\epsilon),
\end{align*}
and, if $C_2=C_{1}^{\perp}$, then $h_{C_1/C_1^{\perp}}$ is symmetric about $\epsilon=1/2$.
\end{lem}

\begin{proof}
For a fixed index set $\cE$, let $\overline{\cE}=[n]\setminus\cE$.
Let $C|_{\overline{\cE}}$ denote the code obtained by puncturing $C$ on $\cE$, which is the projection of $C$ onto the coordinates in $\overline{\cE}$.
The kernel of this projection, restricted to $C$, is $C(\cE)$, so
\begin{align*}
\dim C|_{\overline{\cE}}=\dim C-\dim C(\cE).
\end{align*}
A vector $v\in\bbF_2^{\overline{\cE}}$ lies in $(C|_{\overline{\cE}})^\perp$ if and only if its extension by zeros on $\cE$ lies in $C^\perp$.
The dimension of the orthogonal complement therefore gives
\begin{align}\label{eq:final1-shortening}
\dim C^\perp(\overline{\cE})
=|\overline{\cE}|-\dim C+\dim C(\cE).
\end{align}
Applying this identity to $C_1$ and $C_2$ and subtracting yields
\begin{align*}
&\dim C_1(\cE)-\dim C_2(\cE)
+\dim C_2^\perp(\overline{\cE})
-\dim C_1^\perp(\overline{\cE})\\
&\hspace{2cm}=\dim C_1-\dim C_2.
\end{align*}
By \eqref{eq:bec-entropy-difference}, the first difference equals $H(M\mid E=\cE,Y)$.
The second is the corresponding quantity for the dual nested pair on the complementary set.
If $E$ has erasure probabilities $\bm\epsilon$, then its complement has erasure probabilities $\mone-\bm\epsilon$.
Averaging proves \eqref{eq:bec-folding}.

On the diagonal, differentiation gives
\begin{align*}
A_{C_1/C_2}'(\epsilon)
=A_{C_2^\perp/C_1^\perp}'(1-\epsilon).
\end{align*}
Dividing by $n$ proves the EXIT identity.
When $C_2=C_1^\perp$, the two nested pairs coincide, which proves the claimed symmetry.
\end{proof}

\subsection{Coset Coding on the BSC}
For a linear code $C\subseteq \bbF_{2}^n$, consider a uniformly chosen codeword $U\in C$ which is transmitted through a BSC $W$ and let $Y$ be the channel output. Consider the vector $\bp=[p_1,\dots,p_n]$ where $p_j \in [0,\frac{1}{2}]$ denotes the probability of bit flip for the $j$th channel input.
Then, for $u\in \{0,1\}$,
\begin{align*}
    \Pr(Y_j=u+1\mid U_j=u)=p_{j}.
\end{align*}
For vector $\bp \in [0,\frac{1}{2}]^n$, define the conditional entropy
\begin{align*}
    G_{C}(\bp)\coloneqq H(U\mid Y) 
\end{align*}
and the shorthand $G_{C}(p) \coloneqq G_{C}(p,\dots,p)$.
For a random extrinsic observation $Y_{\sim j}$, the random extrinsic probability is
\begin{align*}
    Q_{j}\coloneqq \Pr(U_{j}=1\mid Y_{\sim j}).
\end{align*}

\begin{defn}[{\cite{Measson-it09}}] \label{def:gexit_bsc}
Let \(U\) be uniformly distributed over a binary linear code \(C\subseteq \bbF_2^n\) and $Y$ be the output of a BSC with input $U$ and error probability \(p\).
Then, the normalized GEXIT function is defined to be
\begin{align*}
    g_{C} (p) \coloneqq h_{2}'(p)- \frac{1}{n} \sum_{j=1}^{n} \mathbb{E} [(1-2Q_j)h_{2}'(p\cnop Q_j)],
\end{align*}
where \(p\cnop q\coloneqq p(1-q)+(1-p)q\).
\end{defn}

\begin{theorem}[{\cite{Measson-it09}}] \label{thm:gexit_bsc_area}
Under the setup in Definition~\ref{def:gexit_bsc}, we have
\begin{align*}
    g_C(p)= \frac{1}{n}G'_{C}(p).
\end{align*}
It follows that
\[
        \int_0^{\frac{1}{2}} g_C(p)\,dp
        =
        \frac{G_C \big(\frac{1}{2}\big)-G_C(0)}{n}
        =
        \frac{\dim C}{n}.
\]
\end{theorem}

\begin{proof}
This was established in~\cite{Measson-it09}.
For completeness, we include a proof in Appendix~\ref{app:known-proofs}.
\end{proof}

We now extend the BEC analysis of coset coding to the BSC.
For nested linear codes $C_2\subseteq C_1\subseteq\bbF_2^n$, let $U$ be uniform on $C_1$ with coset message $M\in T$ as in \eqref{eq:coset decomposition}, and let $Y$ be the output when coordinate $j$ is transmitted through a BSC with crossover probability $p_j$.
Define the coset entropy function
\begin{align*}
    G_{C_1/C_2}(\bp)\coloneqq H(M\mid Y).
\end{align*}
Let $V$ be uniform on $C_2$ and let $Y'$ be its output over the same channels. Lemma~\ref{lem:entropy-difference} then gives
\begin{align}
     G_{C_1/C_2}(\bp) & = H(U\mid Y)-H(V\mid Y') \nonumber \\
     & = G_{C_1}(\bp) -G_{C_2}(\bp) \label{eq:coset_bsc}
\end{align}
with shorthand $G_{C_1/C_2}(p)\coloneqq G_{C_1/C_2}(p,\ldots,p)$.

\begin{defn} \label{def:coset_gexit_bsc}
For the coset coding problem on the BSC with $C_2\subseteq C_{1}$, the normalized coset GEXIT function is defined by
\begin{align*}
    g_{C_1/C_2}(p)
    & \coloneqq g_{C_1}(p) - g_{C_2} (p).
\end{align*}
\end{defn}

\begin{theorem} \label{thm:coset_gexit_bsc_area}
Under the setup in Definition~\ref{def:coset_gexit_bsc}, we have
\[
  g_{C_1/C_2}(p) = \frac{1}{n} \frac{d}{dp} G_{C_1/C_2} (p). 
\]
Moreover,
\[
        \int_0^\frac{1}{2} g_{C_1/C_2} (p)\,dp
        =
        \frac{\dim C_1 - \dim C_2}{n}.
\]
\end{theorem}

\begin{proof}
\eqref{eq:coset_bsc} and Theorem~\ref{thm:gexit_bsc_area} give, for $0<p<1/2$,
\begin{align*}
\frac{1}{n}G_{C_1/C_2}'(p)
=\frac{1}{n}\bigl(G_{C_1}'(p)-G_{C_2}'(p)\bigr)
=g_{C_1}(p)-g_{C_2}(p),
\end{align*}
which is the coset GEXIT function in Definition~\ref{def:coset_gexit_bsc}.
The two single-code GEXIT functions are nonnegative and integrable by the proof of Theorem~\ref{thm:gexit_bsc_area}.
Their difference is therefore integrable as well.
At $p=0$, $Y=U$ determines the message.
At $p=1/2$, $Y$ is independent of the uniform message.
Hence,
\begin{align*}
G_{C_1/C_2}(0)=0,\qquad
G_{C_1/C_2}(1/2)=H(M)=\dim C_1-\dim C_2.
\end{align*}
Applying the fundamental theorem of calculus on $[\delta,1/2]$ and taking $\delta\downarrow0$ gives the claimed area.
\end{proof}

It is useful to express the BSC area theorem in terms of the channel-noise entropy rather than the crossover probability, because this is the parameterization used for Pauli channels in Section~\ref{sec:stabilizer}. Let
\begin{align*}
    a\coloneqq h_2(p), \qquad 0\leq a\leq1,
\end{align*}
and let $p=h_2^{-1}(a)$ denote the inverse on $[0,1/2]$. We define the entropy-parameterized coset GEXIT function by
\begin{align}\label{eq:bsc-entropy-gexit}
    \gamma_{C_1/C_2}^{\mathrm{BSC}}(a)
    \coloneqq \frac{1}{n}\frac{d}{da}
    G_{C_1/C_2}\bigl(h_2^{-1}(a)\bigr)
    =\frac{g_{C_1/C_2}(p)}{h_2'(p)},
\end{align}
where the second equality holds for $0<p<1/2$. A change of variables in the coset GEXIT area theorem gives
\begin{align*}
    \int_0^1\gamma_{C_1/C_2}^{\mathrm{BSC}}(a)\,da
    =\frac{\dim C_1-\dim C_2}{n}.
\end{align*}

\begin{lem}[Pointwise bound for BSC coset GEXIT]
\label{lem:final1-bsc-coset-pointwise}
For $C_2\subseteq C_1\subseteq\bbF_2^n$ and $0<p<1/2$,
\begin{align*}
    0\leq g_{C_1/C_2}(p)\leq h_2'(p).
\end{align*}
Equivalently, the entropy-parameterized function in \eqref{eq:bsc-entropy-gexit} satisfies
\begin{align*}
    0\leq\gamma_{C_1/C_2}^{\mathrm{BSC}}(a)\leq1,
    \qquad 0<a<1.
\end{align*}
\end{lem}
\begin{proof}
Fix a coordinate $j$ and write $Z\coloneqq Y_{\sim j}$. In addition to $Q_j=\Pr(U_j=1\mid Z)$, define
\begin{align*}
    Q_{j,M}\coloneqq\Pr(U_j=1\mid M,Z).
\end{align*}
The entropy chain rule gives
\begin{align*}
H(M\mid Y)
=H(M\mid Z)+H(Y_j\mid M,Z)-H(Y_j\mid Z).
\end{align*}
The first term does not depend on $p_j$. The other two terms satisfy
\begin{align*}
H(Y_j\mid M,Z)&=\mathbb E\bigl[h_2(p_j\cnop Q_{j,M})\bigr],\\
H(Y_j\mid Z)&=\mathbb E\bigl[h_2(p_j\cnop Q_j)\bigr].
\end{align*}
For $0<p<1/2$, define
\begin{align*}
    \phi_p(q)\coloneqq(1-2q)h_2'(p\cnop q).
\end{align*}
Since $1-2q$ and $h_2'(p\cnop q)$ have the same sign, and $p\le p\cnop q\le 1-p$ implies $|h_2'(p\cnop q)|\le h_2'(p)$, we have
\begin{align*}
    0\leq\phi_p(q)\leq h_2'(p), \qquad q\in[0,1].
\end{align*}
Differentiating with respect to $p_j$ therefore gives
\begin{align*}
\frac{\partial}{\partial p_j}G_{C_1/C_2}(\bp)
=\mathbb E[\phi_{p_j}(Q_{j,M})]-\mathbb E[\phi_{p_j}(Q_j)]
\leq h_2'(p_j).
\end{align*}
For the lower bound, increasing $p_j$ produces a degraded observation in coordinate $j$. Indeed, when $p_j\leq p_j'<1/2$, the output with crossover probability $p_j'$ is obtained by independently flipping the output with probability
\begin{align*}
    \delta=\frac{p_j'-p_j}{1-2p_j}.
\end{align*}
Thus, $M\to Y\to\tilde Y$ forms a Markov chain, where $\tilde Y$ is the observation with crossover probability $p_j'$ in coordinate $j$, and data processing shows that $H(M\mid Y)$ is nondecreasing in $p_j$, and hence
\begin{align*}
    0\leq\frac{\partial}{\partial p_j}G_{C_1/C_2}(\bp)
    \leq h_2'(p_j).
\end{align*}
Setting all crossover probabilities equal to $p$, summing over $j$, and dividing by $n$ proves the bound on $g_{C_1/C_2}$. The bound in the entropy parameter follows from \eqref{eq:bsc-entropy-gexit}.
\end{proof}

\section{Area Theorems for Stabilizer Codes}\label{sec:stabilizer}

\subsection{Stabilizer Codes and the Logical Quotient}

A Pauli operator on $n$ qubits is represented, up to phase, by a vector $w=(w_1,\ldots,w_n)\in(\bbF_2^2)^n$, where $w_j=(w_{j,x},w_{j,z})$ identifies the Pauli operator $X^{w_{j,x}}Z^{w_{j,z}}$ on qubit $j$. For $w,w'\in(\bbF_2^2)^n$, the symplectic inner product is defined by
\begin{align*}
  \langle w,w'\rangle_{\symp}\coloneqq\sum_{j=1}^n(w_{j,x}w'_{j,z}+w_{j,z}w'_{j,x}),
\end{align*}
and two Pauli operators commute if and only if their symplectic inner product vanishes.

An $[[n,k]]$ stabilizer code is specified by an additive code $\cS\subseteq(\bbF_2^2)^n$ of dimension $n-k$ over $\bbF_2$.
Since the associated stabilizer operators form a commutative subgroup of the Pauli operators, it follows that $\cS\subseteq\cS^{\perp}$, where
\begin{align*}
    \cS^{\perp}\coloneqq \{a\in(\bbF_2^2)^n\colon \langle a,c\rangle_{\symp}=0,\;\forall c\in\cS\}
\end{align*}
is the dual with respect to the symplectic inner product~\cite{Calderbank-it98*2}.
The corresponding nested pair of additive codes and their quotient are defined by
\begin{align}\label{eq:stabilizer-nesting}
    C_2\coloneqq\cS, \qquad C_1\coloneqq\cS^{\perp}, \qquad
    \cQ\coloneqq C_1/C_2,
\end{align}
so that $\cQ$ is the binary representation of the logical Pauli group. Since $\dim\cS^\perp=2n-\dim\cS=n+k$, we have $|\cQ|=2^{2k}$.
An essential structural point is that symplectic duality exchanges the two codes in the nested pair: $C_2^\perp=C_1$ and $C_1^\perp=C_2$.
Consequently, the quotient associated with the dual pair is the same logical quotient,
\begin{align}\label{eq:self-dual-quotient} 
C_{2}^{\perp}/C_{1}^{\perp}=\cS^{\perp}/\cS=C_1/C_2.
\end{align}

\subsection{CSS Codes and the Logical Quotient}\label{sec:css-setup}

CSS codes~\cite{Calderbank-physreva96,Steane-prl96} are stabilizer codes characterized by a pair of binary classical codes $\cx$ and $\cz$ such that
\begin{align}\label{eq:css-code-containment}
    \cz^{\perp}\subseteq \cx,
\end{align}
or equivalently $\cx^{\perp}\subseteq \cz$.
Alternatively, they are characterized by parity-check matrices $H_{X}\in \mathbb{F}_{2}^{r_{X}\times n}$ and $H_{Z}\in \mathbb{F}_{2}^{r_{Z}\times n}$ satisfying $ H_{X}H_{Z}^{T}=\mzero$ and
\begin{align}
    \cx=\ker H_{X},\quad \cz=\ker H_{Z}.
\end{align}

The row spaces of $H_X$ and $H_Z$ are $C_X^\perp$ and $C_Z^\perp$, respectively.
The rows of $H_X$ specify the $X$-type stabilizers, while the rows of $H_Z$ specify the $Z$-type stabilizers.
Hence, $H_Z$ determines the syndrome of an $X$ error and $H_X$ determines the syndrome of a $Z$ error.
The condition $H_XH_Z^T=0$ ensures that these stabilizers commute.

For binary vectors $x,z\in\bbF_2^n$, write $X^xZ^z$ for the Pauli operator whose bit- and phase-error components are $x$ and $z$.
Using the $(x,z)$ ordering, the stabilizer and the Pauli operators that commute with every stabilizer are represented by
\begin{align*}
    \cS=C_X^\perp\times C_Z^\perp,
    \qquad
    \cS^\perp=C_Z\times C_X.
\end{align*}
The logical Pauli quotient therefore factors as
\begin{align}\label{eq:css-quotient}
    \cQ=\cS^\perp/\cS
    \cong
    C_Z/C_X^\perp\times C_X/C_Z^\perp.
\end{align}
The first factor describes logical classes of $X$ corrections and the second describes logical classes of $Z$ corrections.
We denote these classes by $M_X$ and $M_Z$, respectively.
Each quotient has binary dimension
\begin{align*}
    \dim C_X+\dim C_Z-n=k,
\end{align*}
where $k$ is the number of encoded qubits.
Thus, the joint logical correction class has $2k$ bits of entropy when it is uniform.

The CSS erasure EXIT area theorem is established below in Section~\ref{sec:css-erasure}, and the GEXIT area theorem for independent bit- and phase-flip noise is given in Appendix~\ref{sec:css}.
In both cases, the logical correction problem separates into the two binary coset-decoding problems $C_Z/C_X^\perp$ and $C_X/C_Z^\perp$, and the CSS function is the average of the corresponding classical coset functions.

\subsection{Pauli Channels}\label{sec:pauli-channels}

We consider three families of Pauli channels, each described by the distribution of a Pauli-error vector $\be\in(\bbF_2^2)^n$ with independent elements. The channel parameters are identical across coordinates unless stated otherwise.

The quantum erasure channel with erasure probability $\epsilon\in[0,1]$ erases each qubit independently with probability $\epsilon$ and reveals the erasure set $E$ to the decoder.
Replacing each erased qubit by the maximally mixed state is equivalent to applying a uniformly random Pauli error, so $\be_j$ is uniform on $\bbF_2^2$ when $j\in E$ and $\be_j=0$ otherwise.
For a fixed erasure pattern, each erased qubit contributes two bits of noise entropy and each unerased qubit contributes none. Hence, the conditional entropy of the noise given the erasure pattern is $H(\be\mid E)=2n\epsilon$, or $2\epsilon$ bits per qubit.

The qubit depolarizing channel with parameter $p\in[0,3/4]$ maps an input quantum state $\rho$ to
\begin{align*}
    \cN_p(\rho)
    =(1-p)\rho+\frac{p}{3}
    \left(X\rho X+Y\rho Y+Z\rho Z\right),
\end{align*}
so that $p$ is the total probability of a nonidentity error.
In the Pauli representation, $\Pr(\be_j=0)=1-p$ and $\Pr(\be_j=u)=p/3$ for each $u\in\bbF_2^2\setminus\{0\}$.
There is no side information, and the noise entropy is $h_4(p)\coloneqq h_2(p)+p\log_2 3$ bits per qubit, where $h_2$ is the binary entropy function.
This increases from $0$ to $2$ as $p$ ranges over $[0,3/4]$.

Under independent bit- and phase-flip noise with flip probability $p\in[0,1/2]$, the components of $\be_j=(x_j,z_j)$ are independent with $\Pr(x_j=1)=\Pr(z_j=1)=p$, and the noise entropy is $2h_2(p)$ bits per qubit.
All results in this paper concern the code-capacity noise model, in which Pauli errors act on the data qubits and the syndrome is measured without errors.

\subsection{Syndrome Decoding is Coset Decoding}\label{sec:syndrome-coset}

In quantum error correction, the decoder cannot observe the channel output directly and has access only to the syndrome.
The next lemma shows that the resulting correction problem has the same coset entropy as the classical additive observation $U+\be$.

Let $\be\in(\bbF_2^2)^n$ be the Pauli-error vector, let $S$ denote its measured syndrome, and write $s$ for a realization of $S$.
For each $s$, choose a Pauli correction $\cD(s)$ with syndrome $s$. Since $\be$ and $\cD(S)$ have the same syndrome, the residual error $\br\coloneqq\be+\cD(S)$ belongs to $C_1$. The residual logical correction class is $M\coloneqq\br+C_2\in C_1/C_2$. Here, $M$ denotes a coset rather than a coset representative (in contrast to Section~\ref{sec:background}) but identifying each coset with a unique representative in $T$ does not change the entropy. For a fixed syndrome, a different choice of $\cD(s)$ translates every element of $C_1/C_2$ by the same quotient element, thereby permuting the logical correction classes without changing their conditional entropy.

\begin{lem}\label{lem:syndrome}
Let $U$ be uniform on $C_1$ and independent of $\be$, and let $M_U\coloneqq U+C_2\in C_1/C_2$ and $Y=U+\be$.
Then, for any distribution of $\be$, we have $H(M\mid S)=H(M_U\mid Y)$.
For the quantum erasure channel, if $U$ is independent of $(\be,E)$, we also have $H(M\mid S,E)=H(M_U\mid Y,E)$.
\end{lem}

\begin{proof}
Fix an output $y$ of positive probability and let $s$ be its syndrome.
Every element of $C_1$ has zero syndrome, so $y=U+\be$ and $\be$ have the same syndrome.
For any $u\in C_1$, uniformity of $U$ gives
\begin{align*}
\Pr(M_U=u+C_2\mid Y=y)
=\frac{\Pr(\be\in y+u+C_2)}{\Pr(\be\in y+C_1)}.
\end{align*}
Indeed, summing $\Pr(U=v,Y=y)=|C_1|^{-1}\Pr(\be=y+v)$ over $v\in u+C_2$ gives the numerator before normalization.
Summing over all $v\in C_1$ gives the denominator.

The event $\be\in y+C_1$ is exactly $S=s$.
Also, by the definition of the residual logical correction class,
\begin{align*}
\{\be\in y+u+C_2\}
=\{M=y+\cD(s)+u+C_2,\ S=s\}.
\end{align*}
It follows that
\begin{align*}
\Pr(M_U=u+C_2\mid Y=y)
=\Pr(M=y+\cD(s)+u+C_2\mid S=s).
\end{align*}
Since $y+\cD(s)\in C_1$, this is a translation of the quotient $C_1/C_2$.
It permutes the conditional probabilities and preserves their entropy.
Thus, $H(M_U\mid Y=y)=H(M\mid S=s)$.
Averaging over $y$ proves the first identity.

For erasures, condition first on $E=\cE$.
The same calculation applies with every error probability conditioned on $\cE$.
The uniform input remains independent of the error and erasure set.
Averaging over $y$ and $\cE$ proves $H(M_U\mid Y,E)=H(M\mid S,E)$.
\end{proof}

\subsection{Stabilizer Codes on the Quantum Erasure Channel}

Suppose coordinate $j$ is erased independently with probability $\epsilon_j$. Define
\begin{align*}
    A_{\cQ}(\bm\epsilon)
    \coloneqq H(M\mid S,E),
\end{align*}
and write $A_{\cQ}(\epsilon)\coloneqq A_{\cQ}(\epsilon,\ldots,\epsilon)$. The normalized stabilizer EXIT function is
\begin{align*}
    \gamma_{\cQ}(\epsilon)
    \coloneqq
    \frac{1}{2n}\frac{d}{d\epsilon}A_{\cQ}(\epsilon).
\end{align*}

\begin{theorem}[Stabilizer EXIT area theorem for the quantum erasure channel]
\label{thm:stabilizer-erasure-area}
For an $[[n,k]]$ stabilizer code, $A_{\cQ}(\epsilon)$ is nondecreasing and satisfies
\begin{align*}
    A_{\cQ}(0)=0,
    \qquad
    A_{\cQ}(1)=2k.
\end{align*}
For $0\leq\epsilon\leq\epsilon'\leq1$,
\begin{align}\label{eq:stabilizer-erasure-slope}
    0\leq A_{\cQ}(\epsilon')-A_{\cQ}(\epsilon)
    \leq 2n(\epsilon'-\epsilon).
\end{align}
Consequently,
\begin{align*}
    0\leq\gamma_{\cQ}(\epsilon)\leq1,
    \qquad
    \int_0^1\gamma_{\cQ}(\epsilon)\,d\epsilon=\frac{k}{n}.
\end{align*}
\end{theorem}
\begin{proof}
Let $U$ be uniform on $C_1$ and independent of $(\be,E)$, and let $Y=U+\be$ and $M_U=U+C_2$. By Lemma~\ref{lem:syndrome}, it suffices to consider $H(M_U\mid Y,E)$, and the argument parallels the proof of Theorem~\ref{thm:exit_area} for the BEC. Fix a coordinate $j$ and let $Z=(E_{\sim j},Y_{\sim j})$. When coordinate $j$ is not erased, its observation determines $U_j$. When it is erased, its observation is independent of $U_j$. It follows that
\begin{align*}
A_{\cQ}(\bm\epsilon)
&=(1-\epsilon_j)H(M_U\mid Z,U_j)
+\epsilon_jH(M_U\mid Z).
\end{align*}
The two conditional entropies on the right do not depend on $\epsilon_j$. Therefore,
\begin{align*}
\frac{\partial A_{\cQ}(\bm\epsilon)}{\partial\epsilon_j}
&=H(M_U\mid Z)-H(M_U\mid Z,U_j)\nonumber\\
&=I(M_U;U_j\mid Z).
\end{align*}
Since $U_j\in\bbF_2^2$, this derivative lies in $[0,2]$. On the diagonal, summing over the coordinates gives
\begin{align*}
    0\leq\frac{d}{d\epsilon}A_{\cQ}(\epsilon)\leq2n,
\end{align*}
which proves monotonicity and \eqref{eq:stabilizer-erasure-slope}.

The entropy $A_{\cQ}(\epsilon)$ is a finite weighted sum over erasure sets and is therefore polynomial in $\epsilon$. At $\epsilon=0$, the error is zero and the logical correction class is determined. At $\epsilon=1$, the Pauli error is uniform on $(\bbF_2^2)^n$. Conditioned on a syndrome, the logical correction class is then uniform on $\cQ$. Hence,
\begin{align*}
    A_{\cQ}(0)=0,
    \qquad
    A_{\cQ}(1)=\log_2|\cQ|=2k.
\end{align*}
Dividing the derivative bound by $2n$ gives $0\leq\gamma_{\cQ}(\epsilon)\leq1$. The fundamental theorem of calculus gives
\begin{align*}
    \int_0^1\gamma_{\cQ}(\epsilon)\,d\epsilon
    =\frac{A_{\cQ}(1)-A_{\cQ}(0)}{2n}
    =\frac{k}{n}.
\end{align*}
\end{proof}

\subsection{CSS Codes on the Quantum Erasure Channel}\label{sec:css-erasure}

Let $E$ denote the set of erased coordinates.
Conditioned on $E$, the Pauli error on the erased qubits can be represented by independent binary vectors $x$ and $z$ that are uniform on the vectors supported within $E$.
The decoder knows $E$ and obtains the two syndrome components
\begin{align*}
    S_Z=H_Zx,
    \qquad
    S_X=H_Xz.
\end{align*}
For a fixed value of $S_Z$, any two compatible $X$-error vectors differ by an element of $C_Z$.
Two such errors have the same logical effect when their difference belongs to the $X$-stabilizer space $C_X^\perp$.
The remaining $X$-correction uncertainty is therefore described by the quotient $C_Z/C_X^\perp$.
Likewise, the $Z$-correction uncertainty is described by $C_X/C_Z^\perp$.

Since $x$ and $z$ are independent conditioned on $E$, and each syndrome component depends on only one of them, the two logical correction classes remain conditionally independent.
We define the CSS correction-class entropy by
\begin{align*}
    A_{\mathrm{CSS}(C_X,C_Z)}(\bm\epsilon)
    &\coloneqq H(M_X,M_Z\mid S_Z,S_X,E)\\
    &=H(M_X\mid S_Z,E)+H(M_Z\mid S_X,E).
\end{align*}
As shown in the proof of Theorem~\ref{thm:css-erasure-exit}, these two terms equal the classical erasure coset entropies, which gives
\begin{align}\label{eq:css-erasure-decomposition}
    A_{\mathrm{CSS}(C_X,C_Z)}(\bm\epsilon)
    =A_{C_Z/C_X^\perp}(\bm\epsilon)
    +A_{C_X/C_Z^\perp}(\bm\epsilon).
\end{align}
We use $A_{\mathrm{CSS}(C_X,C_Z)}(\epsilon)$ when all coordinates have the same erasure probability and define
\begin{align*}
    h_{\mathrm{CSS}(C_X,C_Z)}(\epsilon)
    \coloneqq
    \frac{1}{2n}\frac{d}{d\epsilon}
    A_{\mathrm{CSS}(C_X,C_Z)}(\epsilon).
\end{align*}
The factor $1/2$ accounts for the two binary components of each Pauli error. Since $(M_X,M_Z)$ and $(S_Z,S_X)$ are the logical correction class and syndrome of the CSS code viewed as a stabilizer code, $A_{\mathrm{CSS}(C_X,C_Z)}=A_{\cQ}$ and $h_{\mathrm{CSS}(C_X,C_Z)}=\gamma_{\cQ}$. Thus, Theorem~\ref{thm:css-erasure-exit} refines Theorem~\ref{thm:stabilizer-erasure-area} by splitting the stabilizer EXIT function into its $X$ and $Z$ components.

\begin{theorem}[CSS erasure EXIT area theorem]\label{thm:css-erasure-exit}
For a CSS code defined by $C_X$ and $C_Z$, we have
$$ 
    h_{\mathrm{CSS}(C_X,C_Z)}(\epsilon)
    =
    \frac{1}{2}
    \left(
        h_{C_Z/C_X^\perp}(\epsilon)
        +
        h_{C_X/C_Z^\perp}(\epsilon)
    \right).
$$
Since the code has $k=\dim C_X+\dim C_Z-n$ encoded qubits,
$$
    \int_0^1 h_{\mathrm{CSS}(C_X,C_Z)}(\epsilon)\,d\epsilon
    =
    \frac{k}{n}.
$$
\end{theorem}

\begin{proof}
Fix an erasure set $\cE$.
The bit- and phase-error vectors $x$ and $z$ are independent and uniform on the binary vectors supported on $\cE$.
Let $S_Z=H_Zx$ and $S_X=H_Xz$ be the two syndrome components.
Conditioning on these syndromes preserves independence between $x$ and $z$, since each constraint involves only its own error vector.
For a syndrome pair of positive probability, choose supported solutions $x_0$ and $z_0$.
The compatible errors are uniform on
\begin{align*}
x_0+C_Z(\cE),\qquad z_0+C_X(\cE).
\end{align*}
Two compatible $x$ errors have the same logical action exactly when their difference belongs to $C_X^\perp(\cE)$.
The corresponding condition for $z$ errors is membership in $C_Z^\perp(\cE)$.
Thus, the conditional logical-class entropy is
\begin{align*}
H(M\mid S,E=\cE)
&=\dim C_Z(\cE)-\dim C_X^\perp(\cE)\\
&\quad+\dim C_X(\cE)-\dim C_Z^\perp(\cE).
\end{align*}
The supported corrections used for this count may depend on $\cE$.
For each fixed syndrome and erasure set, replacing them by the correction $\cD(s)$ from Section~\ref{sec:syndrome-coset} only translates the logical classes and does not change the entropy.

Averaging over $\cE$ gives
\begin{align*}
A_{\mathrm{CSS}(C_X,C_Z)}(\bm\epsilon)
=A_{C_Z/C_X^\perp}(\bm\epsilon)+A_{C_X/C_Z^\perp}(\bm\epsilon).
\end{align*}
Differentiating on the diagonal and dividing by $2n$ shows that $h_{\mathrm{CSS}(C_X,C_Z)}$ is the average of the two coset EXIT functions.
Each quotient has binary dimension
$\dim C_X+\dim C_Z-n=k$.
Theorem~\ref{thm:coset_exit_area} therefore gives
\begin{align*}
\int_0^1h_{\mathrm{CSS}(C_X,C_Z)}(\epsilon)\,d\epsilon
=\frac12\left(\frac{k}{n}+\frac{k}{n}\right)=\frac{k}{n}.
\end{align*}
\end{proof}

\subsection{Stabilizer Codes on the Depolarizing Channel}

Recall from Section~\ref{sec:pauli-channels} that the single-qubit depolarizing error has entropy $h_4(p)$, which increases from zero to two bits as $p$ increases from zero to $3/4$.
We therefore parameterize the channel by the normalized Pauli-error entropy
\begin{align*}
    a(p)\coloneqq\frac{h_4(p)}{2},
    \qquad 0\leq a(p)\leq1.
\end{align*}
If coordinate $j$ has parameter $p_j$, write $a_j=a(p_j)$ and $\ba=(a_1,\ldots,a_n)$.
Define
\begin{align*}
    G_{\cQ}(\ba)\coloneqq H(M\mid S),
\end{align*}
and use $G_{\cQ}(a)\coloneqq G_{\cQ}(a,\ldots,a)$ when all coordinates have the same parameter.
The normalized stabilizer GEXIT function is
\begin{align*}
    g_{\cQ}(a)
    \coloneqq
    \frac{1}{2n}\frac{d}{da}G_{\cQ}(a).
\end{align*}

\begin{theorem}[Stabilizer GEXIT area theorem for the depolarizing channel]
\label{thm:stabilizer-depolarizing-area}
For an $[[n,k]]$ stabilizer code, $G_{\cQ}(a)$ is nondecreasing and satisfies
\begin{align*}
    G_{\cQ}(0)=0,
    \qquad
    G_{\cQ}(1)=2k.
\end{align*}
For $0\leq a\leq a'\leq1$,
\begin{align}\label{eq:stabilizer-depolarizing-slope}
    0\leq G_{\cQ}(a')-G_{\cQ}(a)
    \leq2n(a'-a).
\end{align}
Consequently,
\begin{align*}
    0\leq g_{\cQ}(a)\leq1 \quad\text{for } 0<a<1,
    \qquad
    \int_0^1g_{\cQ}(a)\,da=\frac{k}{n}.
\end{align*}
\end{theorem}
\begin{proof}
The boundary values are obtained exactly as in the proof of Theorem~\ref{thm:stabilizer-erasure-area}.
At $a=0$, the Pauli error is zero, and at $a=1$, it is uniform on $(\bbF_2^2)^n$, so the logical correction class is uniform on $\cQ$ given the syndrome.
The only new ingredient is the slope bound.
For depolarizing noise, we can couple two channel parameters by adding independent noise.

Let $U$ be uniform on $C_1$ and independent of $\be$, and let $Y=U+\be$ and $M_U=U+C_2$.
By Lemma~\ref{lem:syndrome}, it suffices to consider $H(M_U\mid Y)$.
Fix a coordinate $j$ and $0\leq p_j<p_j'\leq3/4$, and let $\Delta_j\in\bbF_2^2$ be an independent depolarizing error with parameter
\begin{align*}
    \delta\coloneqq\frac{p_j'-p_j}{1-4p_j/3}\in[0,3/4].
\end{align*}
If $\be_j$ has parameter $p_j$, then $\be_j'\coloneqq \be_j+\Delta_j$ has parameter $p_j'$.
Indeed, its total nonidentity probability is $p_j+\delta-4p_j\delta/3=p_j'$, and symmetry gives equal probabilities for the three nonzero values.

Let $Z\coloneqq Y_{\sim j}$ contain the observations outside coordinate $j$, and write $Y_j=U_j+\be_j$ and $Y_j'=Y_j+\Delta_j$.
The Markov chain
\begin{align*}
    M_U\longrightarrow(Z,Y_j)\longrightarrow(Z,Y_j')
\end{align*}
and the data-processing inequality give
\begin{align*}
0
&\leq H(M_U\mid Z,Y_j')-H(M_U\mid Z,Y_j)\\
&=I(M_U;Y_j\mid Z,Y_j')
=I(M_U;\Delta_j\mid Z,Y_j'),
\end{align*}
where the last equality holds because $Y_j$ and $\Delta_j$ determine each other when $Y_j'$ is known.
Since $M_U$ is a function of $U$, conditioning further on $U$ gives
\begin{align*}
I(M_U;\Delta_j\mid Z,Y_j')
&\leq H(\Delta_j)-H(\Delta_j\mid U,Z,Y_j')\\
&=H(\Delta_j)-H(\Delta_j\mid \be_j')\\
&=I(\Delta_j;\be_j')
=H(\be_j')-H(\be_j).
\end{align*}
Here, $(\Delta_j,\be_j')$ is independent of $(U,Z)$, and $U$ together with $Y_j'$ determines $\be_j'$.
The last equality holds because $H(\be_j'\mid\Delta_j)=H(\be_j)$.
Since $H(\be_j)=h_4(p_j)=2a_j$ and $H(\be_j')=2a_j'$, we obtain
\begin{align}\label{eq:depolarizing-coordinate-slope}
0\leq
G_{\cQ}(a_1,\ldots,a_j',\ldots,a_n)
-G_{\cQ}(\ba)
\leq2(a_j'-a_j).
\end{align}

Changing the coordinates one at a time and summing \eqref{eq:depolarizing-coordinate-slope} proves that $G_{\cQ}(a)$ is nondecreasing and satisfies \eqref{eq:stabilizer-depolarizing-slope}.
For $0<p<3/4$, every error pattern has positive probability, and the joint distribution of $(M_U,Y)$ has fixed support with probabilities that are polynomial in $p$.
Hence, $G_{\cQ}$ is a smooth function of $p$ on $(0,3/4)$.
Since $a'(p)=\frac{1}{2}\log_2\frac{3(1-p)}{p}>0$ on this interval, $G_{\cQ}$ is also a smooth function of $a$ for $0<a<1$.
Dividing the slope bound by $2n$ gives $0\leq g_{\cQ}(a)\leq1$ for $0<a<1$.
Since $G_{\cQ}$ is Lipschitz, the fundamental theorem of calculus gives
\begin{align*}
    \int_0^1g_{\cQ}(a)\,da
    =\frac{G_{\cQ}(1)-G_{\cQ}(0)}{2n}
    =\frac{k}{n}. \tag*{\qedhere}
\end{align*}
\end{proof}

\begin{rem}
Theorems~\ref{thm:stabilizer-erasure-area} and~\ref{thm:stabilizer-depolarizing-area} hold for every stabilizer code, including highly degenerate codes, because the area counts logical correction classes rather than stabilizer-equivalent Pauli representatives.
\end{rem}

\section{Converses from the Area Theorems}\label{sec:converse}
\subsection{Quantum Erasure Channel}

If a sequence of stabilizer codes decodes reliably at erasure probability $\epsilon$, its correction-class entropy becomes asymptotically negligible.
From this, the area theorem and the EXIT slope bound imply that the code rate is at most $1-\epsilon$.
However, the relation between $\cS$ and $\cS^\perp$ also relates the correction-class entropy at $\epsilon$ to its value at the complementary erasure probability $1-\epsilon$.
This additional relation strengthens the rate bound to $1-2\epsilon$.

Fix an erasure set $\cE$ and let $\overline{\cE}\coloneqq[n]\setminus\cE$ denote the set of unerased coordinates. By Lemma~\ref{lem:syndrome}, $H(M\mid S,E=\cE)=H(M_U\mid Y,E=\cE)$ with $U$ uniform on $\cS^\perp$. Since the errors are uniform on the erased coordinates, $Y$ reveals $U$ on $\overline{\cE}$ and nothing on $\cE$, so the argument leading to \eqref{eq:bec-entropy-difference} applies to the additive codes $\cS\subseteq\cS^\perp$ and gives $\dim\cS^\perp(\cE)-\dim\cS(\cE)$. Therefore,
\begin{align*}
    A_{\cQ}(\epsilon)
    =\mathbb E\!\left[\dim\cS^{\perp}(E)-\dim\cS(E)\right].
\end{align*}

\begin{theorem}[Erasure folding identity]\label{thm:folding}
For an $[[n,k]]$ stabilizer code on the quantum erasure channel,
\begin{align*}
    A_{\cQ}(\epsilon)+A_{\cQ}(1-\epsilon)=2k,
    \qquad \epsilon\in[0,1].
\end{align*}
Thus, the stabilizer EXIT function satisfies
$\gamma_{\cQ}(\epsilon)=\gamma_{\cQ}(1-\epsilon)$.
\end{theorem}

\begin{proof}
This is the stabilizer analogue of Lemma~\ref{lem:bec-coset-duality} with $C_2=\cS$ and $C_1=\cS^\perp$; because $(\cS^\perp)^\perp=\cS$, the dual nested pair coincides with the original one.
We first establish the symplectic version of the dimension calculation used in Lemma~\ref{lem:bec-coset-duality}.
For an additive code $C\subseteq(\bbF_2^2)^n$, projection onto $\overline{\cE}$ has kernel $C(\cE)$.
The symplectic inner product is a sum over coordinates and is nondegenerate on the $2|\overline{\cE}|$-dimensional coordinate space.
Orthogonality to the projection is therefore equivalent to membership of the zero extension in $C^\perp$.
It follows that
\begin{align*}
\dim C^\perp(\overline{\cE})
=2|\overline{\cE}|-\dim C+\dim C(\cE).
\end{align*}
Apply this identity to $C=\cS$ and $C=\cS^\perp$.
Since $(\cS^\perp)^\perp=\cS$, subtraction gives
\begin{align*}
&\dim\cS^\perp(\cE)-\dim\cS(\cE)
+\dim\cS^\perp(\overline{\cE})-\dim\cS(\overline{\cE})\\
&\hspace{2cm}=\dim\cS^\perp-\dim\cS=2k.
\end{align*}
By the discussion preceding the theorem, the first difference in this display is the correction-class entropy for erasure set $\cE$, and the second is the same quantity for $\overline{\cE}$.
The complementary set has erasure probability $1-\epsilon$ when the original set has erasure probability $\epsilon$.
Averaging proves
\begin{align*}
A_{\cQ}(\epsilon)+A_{\cQ}(1-\epsilon)=2k.
\end{align*}
Both terms are polynomials in $\epsilon$.
Differentiating gives $A_{\cQ}'(\epsilon)=A_{\cQ}'(1-\epsilon)$, and division by $2n$ proves the EXIT symmetry.
\end{proof}

\begin{theorem}
[Erasure converse]
\label{thm:erasure-converse}
Consider a sequence of $[[n_i,k_i]]$ stabilizer codes with $n_i\to\infty$ and $k_i/n_i\to R$, and let $\cQ_i$ be the logical quotient of the $i$-th code.
For $\epsilon\in[0,1/2]$,
if the logical error probability of the optimal syndrome decoder vanishes at erasure probability $\epsilon$, then
\begin{align*}
    R\leq1-2\epsilon.
\end{align*}
\end{theorem}

\begin{proof}
For the $i$-th code, let $P_e^{(i)}$ be the probability that the optimal syndrome decoder selects an incorrect logical correction class.
Given the syndrome and erasure set, the optimal decoder selects a class of largest conditional probability.
The assumption is $P_e^{(i)}\to0$.

For $k_i>0$, Fano's inequality and the $2^{2k_i}$ possible logical classes give
\begin{align*}
A_{\cQ_i}(\epsilon)
=H(M\mid S,E)
&\le h_2(P_e^{(i)})+P_e^{(i)}\log_2(2^{2k_i}-1)\\
&\le1+2k_iP_e^{(i)}.
\end{align*}
When $k_i=0$, the correction-class entropy is zero, and the last bound remains valid.
Since $0\le k_i/n_i\le1$, it follows that
\begin{align*}
\frac{A_{\cQ_i}(\epsilon)}{2n_i}
\le\frac{1}{2n_i}+\frac{k_i}{n_i}P_e^{(i)}
\longrightarrow0.
\end{align*}
Theorem~\ref{thm:folding} gives
\begin{align*}
\frac{A_{\cQ_i}(1-\epsilon)}{2n_i}
=\frac{k_i}{n_i}-\frac{A_{\cQ_i}(\epsilon)}{2n_i}
\longrightarrow R.
\end{align*}
Absolute continuity and the slope bound in Theorem~\ref{thm:stabilizer-erasure-area} imply, for $0\le\epsilon\le1/2$,
\begin{align*}
\frac{A_{\cQ_i}(1-\epsilon)-A_{\cQ_i}(\epsilon)}{2n_i}
=\int_\epsilon^{1-\epsilon}\gamma_{\cQ_i}(u)\,du
\le1-2\epsilon.
\end{align*}
Taking the limit proves $R\le1-2\epsilon$.
\end{proof}

\begin{rem}
The upper bound in Theorem~\ref{thm:erasure-converse} equals the quantum capacity of the erasure channel for $0\leq\epsilon\leq1/2$~\cite{Bennett-physrevlett97}, so it is a tight asymptotic converse.
Related upper bounds for stabilizer codes on the quantum erasure channel were obtained by Delfosse and Z\'emor through rank arguments~\cite{Delfosse-qic13}.
\end{rem}

\subsection{Depolarizing Channel}

The converse on depolarizing channels requires information about both outputs of the quantum channel. Let $L$ denote the $k$-qubit logical system and let $L'$ be a reference system of the same dimension. We begin with the maximally entangled state
\begin{align*}
    \ket{\Phi}_{L'L}
    \coloneqq 2^{-k/2}\sum_{\ell=1}^{2^k}\ket{\ell}_{L'}\ket{\ell}_L.
\end{align*}
Let $A$ denote a single-qubit channel input, so that the $n$ physical qubits form $A^n$. Let $\{\ket{\bar\ell}\}_{\ell=1}^{2^k}$ be an orthonormal basis of the stabilizer code space, and let $J\coloneqq\sum_{\ell=1}^{2^k}\ket{\bar\ell}_{A^n}\bra{\ell}_L$ be the corresponding isometric encoder. A different choice of basis changes $J$ by a unitary on $L$, which can be moved onto $L'$ and does not change the entropies below. A Stinespring isometry for the depolarizing channel is an isometry $\cV:A\rightarrow B\otimes F$ satisfying
\begin{align*}
    \cN_p(\rho)=\Tr_F\!\left(\cV\rho \cV^\dagger\right)
\end{align*}
for every input quantum state $\rho$ on $A$.
The system $B$ is the channel output and $F$ is the complementary output. Applying the encoder and $\cV^{\otimes n}$ gives the pure state
\begin{align*}
    \ket{\Omega}_{L'B^nF^n}
    =\left(I_{L'}\otimes \cV^{\otimes n}J\right)\ket{\Phi}_{L'L}.
\end{align*}
The complementary channel is
\begin{align*}
    \cN_p^c(\rho)=\Tr_B\!\left(\cV\rho \cV^\dagger\right).
\end{align*}
The complementary channel is defined up to an isometry on $F$, which leaves the entropy quantities below unchanged~\cite{Watrous-2018}.
For the state $\ket{\Omega}$, the coherent information from the reference to the channel output is
\begin{align*}
    I_c(L'\rangle B^n)
    \coloneqq-H(L'\mid B^n)
    =H(B^n)-H(L',B^n).
\end{align*}

\begin{lem}[Correction-class entropy and coherent information]
\label{lem:final1-coherent}
For an $[[n,k]]$ stabilizer code on the depolarizing channel,
\begin{align}\label{eq:depol-coherent-relation}
    G_{\cQ}(a(p))=k-I_c(L'\rangle B^n).
\end{align}
Moreover, define the complementary-output quantity
\begin{align*}
    G_{\cQ}^{F}(p)\coloneqq k-I_c(L'\rangle F^n),
    \qquad
    I_c(L'\rangle F^n)\coloneqq-H(L'\mid F^n).
\end{align*}
Then
\begin{align}\label{eq:depol-output-complement-folding}
    G_{\cQ}(a(p))+G_{\cQ}^{F}(p)=2k.
\end{align}
\end{lem}
\begin{proof}
The Pauli errors can be grouped by their syndrome $S$ and logical correction class $M$. Errors with different syndromes map the code space into mutually orthogonal subspaces. Within a fixed syndrome subspace, distinct logical Pauli operators applied to the encoded half of the maximally entangled state give mutually orthogonal states on $(L',B^n)$. 

More explicitly, after grouping the Pauli errors by $(S,M)$, the reduced state on $(L',B^n)$ is an orthogonal mixture of pure states, one for each pair $(s,m)$, with weights $\Pr(S=s,M=m)$.
Errors with the same $(s,m)$ differ by a stabilizer and hence give the same pure state up to phase, whereas errors with different $(s,m)$ pairs give orthogonal states.
The von Neumann entropy of $(L',B^n)$ therefore equals the Shannon entropy and
\begin{align*}
    H(L',B^n)=H(S)+H(M\mid S).
\end{align*}
After tracing out $L'$, each of these pure states becomes the maximally mixed state $2^{-k}\Pi_s$ on an error-translated code subspace, where $\Pi_s$ is a rank-$2^k$ projector that depends only on the syndrome $s$.
Changing the logical class acts unitarily within the same subspace, while different syndromes give orthogonal subspaces.
Thus the state on $B^n$ is an orthogonal mixture of the states $2^{-k}\Pi_s$ with weights $\Pr(S=s)$, and its entropy is given by
\begin{align*}
    H(B^n)=H(S)+\sum_s\Pr(S=s)k=H(S)+k.
\end{align*}

Therefore,
\begin{align*}
H(L'\mid B^n)
&=H(L',B^n)-H(B^n)\\
&=H(M\mid S)-k
=G_{\cQ}(a(p))-k.
\end{align*}
Using the definition of coherent information proves \eqref{eq:depol-coherent-relation}.
Since $\ket{\Omega}_{L'B^nF^n}$ is pure, $H(L'\mid B^n)=-H(L'\mid F^n)$, and \eqref{eq:depol-output-complement-folding} follows from \eqref{eq:depol-coherent-relation}.
\end{proof}

We next show that part of the complementary output is itself the output of another depolarizing channel.
For $e\in\bbF_2^2$, let $\sigma_e$ denote the corresponding Hermitian Pauli matrix, with $\sigma_0=I$, and write $\lambda_p(e)$ for the depolarizing error probabilities defined in Section~\ref{sec:pauli-channels}.
Let the complementary system $F=F_1F_2$ consist of two qubits, and define the Bell states $\ket{\psi_e}_{F_1F_2}\coloneqq(\sigma_e\otimes I)(\ket{00}+\ket{11})/\sqrt2$, which form an orthonormal basis of $F$.
Now, we choose instead the Stinespring isometry
\begin{align*}
    \cV=\sum_{e\in\bbF_2^2}\sqrt{\lambda_p(e)}\,\sigma_e\otimes\ket{\psi_e}_{F_1F_2}.
\end{align*}

\begin{lem}[Part of the complementary output is depolarizing]
\label{lem:final1-depolarizing-map}
With this choice of $\cV$, discarding $F_2$ from the complementary output gives
\begin{align}\label{eq:final1-environment-depolarizing}
    \Tr_{F_2}\!\left(\cN_p^c(\rho)\right)=\cN_{f(p)}(\rho),
\end{align}
where
\begin{align}\label{eq:final1-depolarizing-involution}
    f(p)=\frac{3-2p-2\sqrt{3p(1-p)}}{4}.
\end{align}
The function $f$ is decreasing on $[0,3/4]$ with $f(1/4)=1/4$, so $f(p)\geq p$ for $p\leq1/4$ and $f(p)\leq p$ for $p\geq1/4$.
\end{lem}
\begin{proof}
The complementary output is
\begin{align*}
    \cN_p^c(\rho)
    =\sum_{e,e'\in\bbF_2^2}\sqrt{\lambda_p(e)\lambda_p(e')}\,
    \Tr(\sigma_{e'}\sigma_e\rho)\,\ket{\psi_e}\bra{\psi_{e'}}.
\end{align*}
Since $\Tr_{F_2}\ket{\psi_e}\bra{\psi_{e'}}=\sigma_e(I/2)\sigma_{e'}$, we obtain
\begin{align*}
    \Tr_{F_2}\!\left(\cN_p^c(\rho)\right)
    =\frac12\sum_{e,e'\in\bbF_2^2}\sqrt{\lambda_p(e)\lambda_p(e')}\,
    \Tr(\sigma_{e'}\sigma_e\rho)\,\sigma_e\sigma_{e'}.
\end{align*}
For $e,e'\in\bbF_2^2$, the Pauli matrices satisfy $\sigma_{e'}\sigma_e=\omega_{e,e'}\sigma_{e+e'}$ and $\sigma_e\sigma_{e'}=\omega_{e,e'}^{-1}\sigma_{e+e'}$ for a phase $\omega_{e,e'}\in\{\pm1,\pm i\}$, so the phases cancel and
\begin{align*}
    \Tr_{F_2}\!\left(\cN_p^c(\rho)\right)
    =\frac12\sum_{g\in\bbF_2^2}\theta_g\Tr(\sigma_g\rho)\sigma_g,
    \qquad
    \theta_g\coloneqq\sum_{e+e'=g}\sqrt{\lambda_p(e)\lambda_p(e')}.
\end{align*}
The coefficient of $\sigma_0=I$ is unchanged because $\theta_0=\sum_e\lambda_p(e)=1$.
For each $g\neq0$, the coefficient of $\sigma_g$ is multiplied by the same factor
\begin{align*}
    \theta(p)=2\sqrt{\frac{(1-p)p}{3}}+\frac{2p}{3}.
\end{align*}
In the Pauli expansion $\rho=\frac12\sum_{g}\Tr(\sigma_g\rho)\sigma_g$, the depolarizing channel $\cN_q$ with $q\in[0,3/4]$ multiplies the coefficient of each nonidentity Pauli matrix $\sigma_g$, $g\neq0$, by $1-4q/3$, because $\cN_q(\rho)=(1-\frac{4q}{3})\rho+\frac{4q}{3}\Tr(\rho)\frac I2$.
Setting $1-4q/3=\theta(p)$ gives $q=f(p)$, which proves \eqref{eq:final1-environment-depolarizing}. Since this calculation is linear in $\rho$ and holds for every operator $\rho$, \eqref{eq:final1-environment-depolarizing} is an identity of linear maps; hence, it also holds when applied qubit by qubit to entangled inputs.
Direct differentiation shows that $f$ is decreasing on $[0,3/4]$, and substitution gives $f(1/4)=1/4$.
\end{proof}

\begin{theorem}[Depolarizing folding inequality]
\label{thm:final1-depolarizing-folding}
For an $[[n,k]]$ stabilizer code and $0\leq p\leq3/4$,
\begin{align}\label{eq:final1-depolarizing-folding}
    G_{\cQ}(a(p))+G_{\cQ}(a(f(p)))\geq2k.
\end{align}
\end{theorem}
\begin{proof}
Discard $F_2$ from each of the $n$ complementary outputs, and denote the remaining system by $F_1^n$. By Lemma~\ref{lem:final1-depolarizing-map}, $F_1^n$ is the output of $\cN_{f(p)}^{\otimes n}$ for the same encoded input. By the data-processing inequality for coherent information, it cannot increase when part of the conditioning system is discarded~\cite{Watrous-2018}. Therefore,
\begin{align*}
I_c(L'\rangle F^n)
&\geq I_c(L'\rangle F_1^n)\\
&=k-G_{\cQ}(a(f(p))),
\end{align*}
where the equality follows from Lemma~\ref{lem:final1-coherent}. Hence,
\begin{align*}
    G_{\cQ}^{F}(p)
    =k-I_c(L'\rangle F^n)
    \leq G_{\cQ}(a(f(p))).
\end{align*}
Combining this inequality with \eqref{eq:depol-output-complement-folding} proves \eqref{eq:final1-depolarizing-folding}.
\end{proof}

\begin{theorem}[Depolarizing converse]
\label{thm:final1-depolarizing-converse}
Consider a sequence of $[[n_i,k_i]]$ stabilizer codes with $n_i\to\infty$ and $k_i/n_i\to R$, and write $\cQ_i$ for the logical quotient of the $i$-th code. Suppose the logical correction-class error probability of the optimal syndrome decoder vanishes on $\cN_p^{\otimes n_i}$ for some $0\leq p\leq3/4$. Then
\begin{align}\label{eq:final1-depolarizing-converse}
    R\leq\left[\frac{h_4(f(p))-h_4(p)}{2}\right]^+,
\end{align}
where $[x]^+\coloneqq\max\{x,0\}$. In particular, $R=0$ for $1/4\leq p\leq3/4$, and no positive-rate stabilizer family has a depolarizing threshold at or above $p=1/4$.
\end{theorem}
\begin{proof}
As in the proof of Theorem~\ref{thm:erasure-converse}, Fano's inequality gives $G_{\cQ_i}(a(p))\leq1+2k_iP_e^{(i)}$, where $P_e^{(i)}\to0$ is the error probability of the optimal syndrome decoder. Hence,
\begin{align}\label{eq:final1-depolarizing-fano}
    \frac{G_{\cQ_i}(a(p))}{2n_i}\longrightarrow0.
\end{align}
If $f(p)\geq p$, the slope bound in Theorem~\ref{thm:stabilizer-depolarizing-area} gives $G_{\cQ_i}(a(f(p)))-G_{\cQ_i}(a(p))\leq2n_i(a(f(p))-a(p))$. If $f(p)<p$, monotonicity gives $G_{\cQ_i}(a(f(p)))-G_{\cQ_i}(a(p))\leq0$. Combining either case with Theorem~\ref{thm:final1-depolarizing-folding} yields
\begin{align*}
2k_i-2G_{\cQ_i}(a(p))
&\leq G_{\cQ_i}(a(f(p)))-G_{\cQ_i}(a(p))\\
&\leq2n_i\bigl[a(f(p))-a(p)\bigr]^+.
\end{align*}
Dividing by $2n_i$, applying \eqref{eq:final1-depolarizing-fano}, and using $a(q)=h_4(q)/2$ proves \eqref{eq:final1-depolarizing-converse}. Since $a$ and $h_4$ are increasing on $[0,3/4]$, the bracket is nonpositive exactly when $f(p)\leq p$, that is, when $p\geq1/4$ by Lemma~\ref{lem:final1-depolarizing-map}.
\end{proof}

\begin{rem}
The bound in Theorem~\ref{thm:final1-depolarizing-converse} is not intended to compete with the best known upper bounds on the quantum capacity of the depolarizing channel for $0<p<1/4$~\cite{Smith-itw08,Sutter-it17,Leditzky-it17,Leditzky-prl18}.
The bound follows from the monotonicity and slope bound in Theorem~\ref{thm:stabilizer-depolarizing-area}, a single folding inequality, and Fano's inequality, and it vanishes exactly at $p=1/4$, where the depolarizing channel becomes antidegradable and its quantum capacity is zero~\cite{Bruss-pra98,Cerf-prl00,Cubitt-jmp08}.

For memoryless Pauli channels, the optimal rate achievable by stabilizer codes with vanishing error is characterized by the regularized coherent information maximized over input states that are maximally mixed on stabilizer code spaces~\cite{Hamada-it05}, and the entanglement fidelity decays exponentially in the blocklength for code families whose asymptotic rate exceeds this value by a fixed positive amount~\cite{Tomamichel-arxiv26}.
In particular, for the depolarizing channel with $1/4\le p\le3/4$, no family of stabilizer codes with positive rate keeps its entanglement fidelity bounded away from zero~\cite{Tomamichel-arxiv26}.
Theorem~\ref{thm:final1-depolarizing-converse} complements these results with an explicit bound for $0<p<1/4$ in terms of $h_4$ and $f$.

As $p\to0$, our bound behaves as $1-h_4(p)/2-p/\ln2$, whereas the hashing rate $1-h_4(p)$ is achievable and agrees with the quantum capacity of the depolarizing channel to leading order in $p$~\cite{Leditzky-prl18}.
Thus, at low noise, our bound exceeds the hashing rate by approximately $h_4(p)/2$.
Since our bound holds for arbitrary stabilizer codes, it also applies to degenerate codes, which can exceed the hashing rate $1-h_4(p)$~\cite{DiVincenzo-pra98,Smith-prl07}.
\end{rem}

\section{Numerical Results}\label{sec:numerical}

In this section, we validate the area theorems numerically on the quantum erasure channel and the depolarizing channel for several classes of codes.
On the quantum erasure channel, we use Monte Carlo simulation over erasure patterns for large planar surface codes and bivariate-bicycle codes.
On the depolarizing channel, we enumerate all Pauli errors of small stabilizer codes and compute the correction-class entropy from the posterior distribution of logical correction classes conditioned on the syndrome. The optimal MAP decoder selects the most probable class under this distribution.
Appendix~\ref{app:numerical-small} gives the corresponding results for the same small codes on the quantum erasure channel and under independent bit- and phase-flip noise.

\subsection{Quantum Erasure Channel}\label{sec:numerical-mc}

\begin{figure}[t]
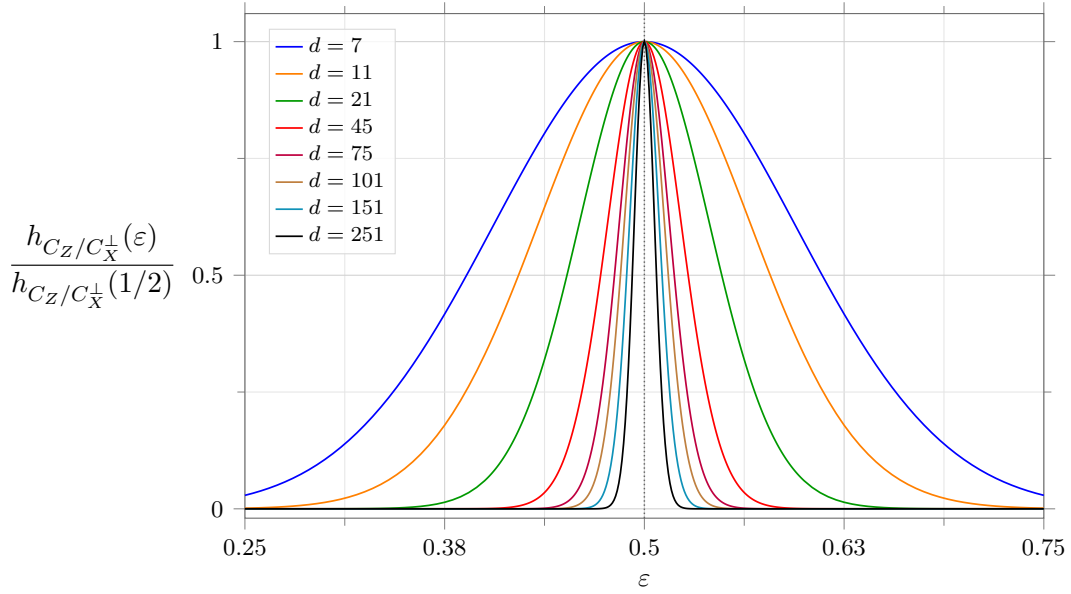

\centering

 \caption{Normalized component EXIT functions $h_{C_Z/C_X^{\perp}}(\epsilon)/h_{C_Z/C_X^{\perp}}(1/2)$ of planar surface codes on the quantum erasure channel, estimated as described in Section~\ref{sec:numerical-mc}. The curves are scaled by $20.10$, $37.25$, $87.54$, $233.6$, $447.0$, $651.0$, $1081$, and $2046$ in legend order. Before this normalization, each component area equals $k/n$.}
\label{fig:mc-exit-surface}
\end{figure}
 \begin{figure}[t]
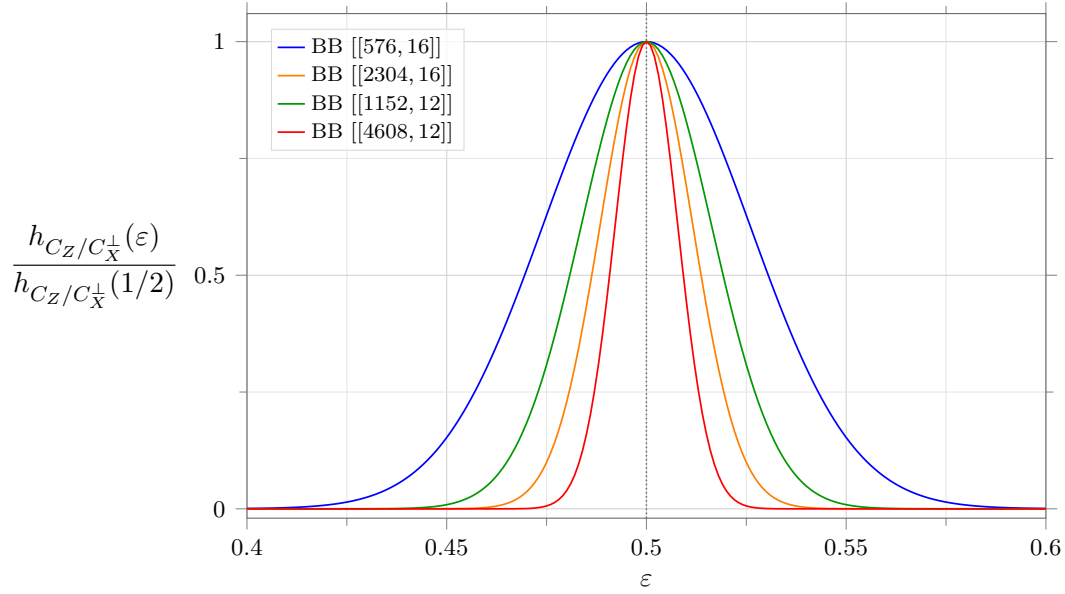

\centering
%
 \caption{Normalized component EXIT functions $h_{C_Z/C_X^{\perp}}(\epsilon)/h_{C_Z/C_X^{\perp}}(1/2)$ of bivariate-bicycle codes on the quantum erasure channel, estimated as described in Section~\ref{sec:numerical-mc}. The curves are scaled by $2.333$, $4.198$, $3.922$, and $7.547$ in legend order. Before this normalization, each component area equals $k/n$.}
\label{fig:mc-exit-bb}
\end{figure}
 \begin{figure}[t]
\centering
%
 \caption{Normalized $X$-component correction-class entropy $A_{C_Z/C_X^{\perp}}(\epsilon)/k$ on the quantum erasure channel for the planar surface codes in Fig.~\ref{fig:mc-exit-surface}. Here $k$ is the number of encoded logical qubits. The largest standard error over $\epsilon\in[0,1]$ is $2.2\times10^{-4}$, which is below the line width.}
\label{fig:mc-class-surface}
\end{figure}
 \begin{figure}[t]
\centering
%
 \caption{Normalized $X$-component correction-class entropy $A_{C_Z/C_X^{\perp}}(\epsilon)/k$ on the quantum erasure channel for the bivariate-bicycle codes in Fig.~\ref{fig:mc-exit-bb}. Here $k$ is the number of encoded logical qubits. The largest standard error over $\epsilon\in[0,1]$ is $7.6\times10^{-5}$, which is below the line width.}
\label{fig:mc-class-bb}
\end{figure}
 
This subsection presents Monte Carlo estimates of the correction-class entropies on the quantum erasure channel for the $[[d^2+(d-1)^2,1,d]]$ planar surface codes with $d\in\{5,7,11,21,45,75,101,151,251\}$ and for the bivariate-bicycle codes $[[576,16]]$, $[[2304,16]]$, $[[1152,12]]$, and $[[4608,12]]$.
Figs.~\ref{fig:mc-exit-surface} and~\ref{fig:mc-exit-bb} show the normalized EXIT functions, Figs.~\ref{fig:mc-class-surface} and~\ref{fig:mc-class-bb} show the normalized correction-class entropies, and Fig.~\ref{fig:mc-decomposition} in Appendix~\ref{app:numerical-surface} shows the correction-class decomposition.
The code constructions, the computation, and the validation are described in Appendix~\ref{app:numerical}.

For a fixed erasure set $\cE\subseteq[n]$, \eqref{eq:bec-entropy} and \eqref{eq:final1-shortening} with $C=C_X$ and the index set $\overline{\cE}$ give, as in the proof of Theorem~\ref{thm:css-erasure-exit},
\begin{align}\label{eq:bb-mc-class-x}
    H(M_X\mid S_Z,E=\cE)
    &=\dim C_Z(\cE)-\dim C_X^\perp(\cE)\nonumber\\
    &=|\cE|-\rank H^{C_Z}_{\cE}-\rank H_X+\rank H^{C_X}_{\overline{\cE}},
\end{align}
and the $Z$-component correction-class entropy $H(M_Z\mid S_X,E=\cE)=\dim C_X(\cE)-\dim C_Z^\perp(\cE)$ follows by exchanging the roles of $X$ and $Z$.
Each sample draws independent random variables $\xi_1,\ldots,\xi_n$ that are uniform on $[0,1]$ and sets $E(\epsilon)\coloneqq\{j\in[n]\colon \xi_j<\epsilon\}$, which is the erasure set of the quantum erasure channel with erasure probability $\epsilon$.
Let $\pi$ be the permutation of $[n]$ that sorts $\xi_1,\ldots,\xi_n$ in increasing order, and let $E_t\coloneqq\{\pi(1),\ldots,\pi(t)\}$ for $t=0,\ldots,n$.
The sets $E_0\subset E_1\subset\cdots\subset E_n$ are nested, $E(\epsilon)=E_t$ with $t=|E(\epsilon)|$, and the ranks in \eqref{eq:bb-mc-class-x} for all $n+1$ sets $E_t$ follow by inserting one column at a time.
For $t=0,\ldots,n$, let $\alpha_t$ be the expectation of \eqref{eq:bb-mc-class-x} when $\cE$ is uniformly distributed over the subsets of $[n]$ of size $t$.
Since $E_t$ is such a subset, the sample average $\hat\alpha_t$ of \eqref{eq:bb-mc-class-x} at $\cE=E_t$ is an unbiased estimate of $\alpha_t$.
Conditioned on $|E|=t$, the erasure set $E$ is uniformly distributed over the sets of size $t$, which gives
\begin{align}
    A_{C_Z/C_X^\perp}(\epsilon)
    &=\sum_{t=0}^{n}\alpha_t\binom{n}{t}\epsilon^t(1-\epsilon)^{n-t},
    \label{eq:bb-mc-bernstein}\\
    h_{C_Z/C_X^\perp}(\epsilon)
    &=\sum_{t=0}^{n-1}\left(\alpha_{t+1}-\alpha_t\right)\binom{n-1}{t}\epsilon^t(1-\epsilon)^{n-1-t},
    \label{eq:bb-mc-bernstein-derivative}
\end{align}
where the second identity follows by differentiating the first and dividing by $n$.
We substitute $\hat\alpha_t$ for $\alpha_t$, so the EXIT function is obtained by exact differentiation of a polynomial in $\epsilon$ and not by difference quotients.

For the surface codes, the interval on which $A_{C_Z/C_X^\perp}(\epsilon)$ increases from $0.1$ to $0.9$ narrows from width $0.3085$ at $d=5$ to $0.0169$ at $d=251$.
This width decreases approximately as $d^{-3/4}$, as discussed in Appendix~\ref{app:numerical-surface}.
For the bivariate-bicycle codes, quadrupling the blocklength at fixed $k$ narrows the corresponding interval for $A_{C_Z/C_X^\perp}(\epsilon)/k$ by a factor of $2.22$ for $k=16$ and $2.08$ for $k=12$.
In both families, the peak value $h_{C_Z/C_X^\perp}(1/2)$ decreases as the transition sharpens, while the area remains $k/n$.
For every code, the estimates satisfy $A_{C_Z/C_X^\perp}(\epsilon)+A_{C_X/C_Z^\perp}(1-\epsilon)=k$ from Lemma~\ref{lem:bec-coset-duality} within $2.94$ standard errors over $\epsilon\in[0,1]$, where the two terms are computed from different erasure sets.
Since $\hat\alpha_0=0$ and $\hat\alpha_n=k$ in every sample, the area under \eqref{eq:bb-mc-bernstein-derivative} equals $k/n$ exactly.

\subsection{Depolarizing Channel}\label{sec:numerical-pauli}

\begin{figure}[t]
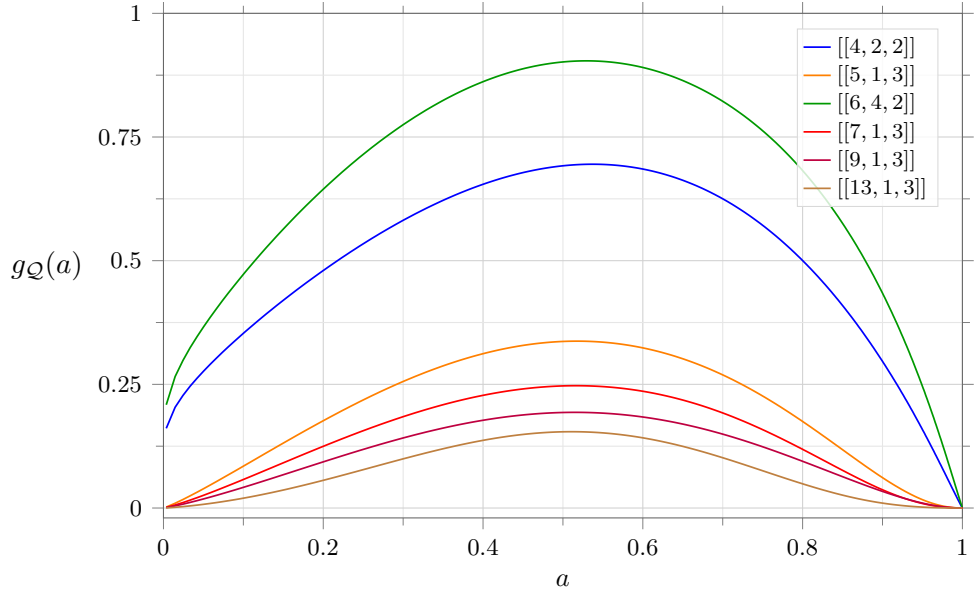

\centering

 \caption{Stabilizer GEXIT function $g_{\cQ}(a)$ of small stabilizer codes on the depolarizing channel with $a=a(p)=h_4(p)/2$, computed as described in Section~\ref{sec:numerical-pauli}. Each area equals $k/n$, and every curve lies in $[0,1]$.}
\label{fig:pauli-exact}
\end{figure}
 
This subsection evaluates the correction-class entropies of small stabilizer codes on the depolarizing channel, and it tests Theorem~\ref{thm:stabilizer-depolarizing-area}, Lemma~\ref{lem:final1-coherent}, and Theorem~\ref{thm:final1-depolarizing-folding}.
The computations use the $[[4,2,2]]$, $[[5,1,3]]$, $[[6,4,2]]$, Steane $[[7,1,3]]$, Shor $[[9,1,3]]$, and $[[13,1,3]]$ planar surface codes.
Grouping all $4^n$ Pauli-error vectors by syndrome, logical correction class, and number of nonzero coordinates gives $\Pr(S=s,M=m)$ as a polynomial in $p$. From this distribution, we compute $G_{\cQ}(a(p))$ and its derivative. The MAP decoder selects the most probable logical correction class conditioned on the syndrome.
For the codes with $n\le9$, we also computed $I_c(L'\rangle B^n)$ directly from the density matrices of the encoded maximally entangled state after the depolarizing channel.
Appendix~\ref{app:numerical-small} gives the code definitions, the computation, and the numerical values.

For all six codes on the depolarizing channel, $g_{\cQ}(a)$ lies in $[0,1]$, as Theorem~\ref{thm:stabilizer-depolarizing-area} states, and its largest value $0.904$ occurs for the $[[6,4,2]]$ code, which has the largest rate.
The maximum of $g_{\cQ}$ occurs at a depolarizing probability between $0.1955$ and $0.2100$ for every code.
For the codes with $n\le9$, $G_{\cQ}(a(p))$ and $k-I_c(L'\rangle B^n)$ agree within $2\times10^{-14}$, which confirms \eqref{eq:depol-coherent-relation} of Lemma~\ref{lem:final1-coherent}, including for the non-CSS $[[5,1,3]]$ code and the degenerate $[[9,1,3]]$ code.
The folding inequality \eqref{eq:final1-depolarizing-folding} holds at every evaluated $p$.
At $p=1/4$, where $f(1/4)=1/4$, it reduces to $G_{\cQ}(a(1/4))\ge k$, and the ratio $G_{\cQ}(a(1/4))/k$ lies between $1.282$ and $1.418$ for the six codes.

\section{Conclusions}\label{sec:conclusion}

This paper formulates EXIT and GEXIT area theorems in terms of the uncertainty that remains about a coset.
For nested binary codes, this gives area identities for coset decoding on the BEC and the BSC.
For stabilizer codes, the corresponding coset is the logical correction class that remains after the syndrome is measured.
Its entropy accounts for stabilizer degeneracy and measures exactly the uncertainty relevant to optimal syndrome decoding.

On the quantum erasure channel, the correction-class entropy defines an EXIT function whose area is the quantum code rate, and the relation between $\cS$ and $\cS^\perp$ connects this entropy at complementary erasure probabilities.
Combining this folding identity with the slope bound yields the tight rate constraint $R\leq1-2\epsilon$ for $0\leq \epsilon\leq 1/2$ .
On the depolarizing channel, the bit- and phase-error components are correlated, so the analysis retains the full four-valued Pauli-error distribution.
Parameterizing the channel by its normalized Pauli-error entropy gives a GEXIT function between zero and one whose area is again the quantum code rate.
Relating the correction-class entropy to coherent information and processing the complementary channel output gives a folded entropy inequality and a rate-dependent converse that reaches zero at $p=1/4$, where the depolarizing channel becomes antidegradable.
For CSS codes on the quantum erasure channel and under independent bit- and phase-flip noise, the correction-class entropy separates into two classical coset entropies, which gives the componentwise interpretation used in the numerical examples.

We believe these results have several broader implications:
\begin{itemize}
    \item The area identities and slope bounds constrain optimal syndrome decoding for every stabilizer code, and the optimal correction-class curves provide a benchmark for practical quantum decoders.
    \item Because the (G)EXIT area is fixed by the rate, the shape of the correction-class curve connects decoding performance directly to rate, which suggests EXIT-based design criteria for coset and CSS codes.
    \item The coset viewpoint connects CSS codes with wiretap codes. Any nested pair $C_2\subseteq C_1$ defines a CSS code with $C_Z=C_1$ and $C_X=C_2^\perp$, and by Lemma~\ref{lem:bec-coset-duality}, reliable decoding of both components on the quantum erasure channel with erasure probability $\epsilon$ requires the coset entropy of $C_1/C_2$ to be nearly zero at $\epsilon$ and nearly maximal at $1-\epsilon$. These are the reliability and secrecy conditions for a binary erasure wiretap channel whose legitimate and eavesdropper channels have erasure probabilities $\epsilon$ and $1-\epsilon$, which suggests a close correspondence between the two design problems.
    \item The correction-class entropy provides a new path toward explaining the gap between optimal decoding and message-passing decoders, such as quantum peeling~\cite{Pech-isit26} and belief-propagation guided decimation~\cite{Yao-isit24}, as well as quantum Maxwell decoding~\cite{Freire-arxiv26}.
\end{itemize}

The results in this paper hold for the code-capacity noise model.
In recent work~\cite{Ruiz-arxiv26}, the conditional entropy of the logical correction class given the syndrome has been used to estimate thresholds of quantum error-correction protocols, including under circuit-level noise, where Pauli errors on the data qubits are replaced by faults in the syndrome-extraction circuit and the syndrome is replaced by detector outcomes.
This suggests that extending the area theorems and folding relations to circuit-level noise may provide a tool for determining whether a decoding method is optimal or suboptimal, and by how much, with respect to achievable encoding rates for fault-tolerant quantum computation.

\section*{Acknowledgments}
The authors thank Alexei Ashikhmin for helpful discussions related to an earlier version of this work.
This research was supported in part by a gift from Western Digital Corporation.
The material is based on work supported by the NSF under Grant 2120757.
Any opinions, findings, and conclusions or recommendations expressed in this material are those of the authors and do not necessarily reflect the views of the NSF.

\section*{AI Statement}

The authors used ChatGPT (GPT-6 Astra and earlier GPT model versions) and Claude (Opus 5.5) to refine the proofs and to improve the clarity and presentation of the manuscript. The main results concerning the connection between coset coding and syndrome decoding were conceived and derived by the authors, who used them to obtain the area theorem for stabilizer codes. The authors later conceived the proof of the erasure converse on their own. Although the authors suspected that the area theorem could be used to obtain a converse for the depolarizing channel, the proof technique based on the complementary channel was suggested by ChatGPT. The authors also used Codex and Claude Code to refine the scripts that generate the numerical results and to draft a detailed discussion of these results. All AI-suggested changes were reviewed and verified by the authors. The authors take full responsibility for the content of the manuscript.
\printbibliography

\appendix

\section{Proofs of Known Area Theorems}\label{app:known-proofs}
This appendix gives the proofs of Theorems~\ref{thm:exit_area} and~\ref{thm:gexit_bsc_area}, which are included for completeness.

\begin{proof}[Proof of Theorem~\ref{thm:exit_area}]
Fix a coordinate $j$ and condition on the erasures and observations outside that coordinate.
The event $j\in E$ is independent of $(U,E_{\sim j},Y_{\sim j})$.
If $j\notin E$, the additional observation determines $U_j$.
If $j\in E$, the additional observation is the erasure symbol.
Consequently,
\begin{align*}
A_C(\bm\epsilon)
&=(1-\epsilon_j)H(U\mid E_{\sim j},Y_{\sim j},U_j)
+\epsilon_j H(U\mid E_{\sim j},Y_{\sim j}).
\end{align*}
Both conditional entropies on the right are independent of $\epsilon_j$.
Differentiating and using that $U_j$ is a function of $U$ gives
\begin{align}\label{eq:partial-derivative-bec-exit}
\frac{\partial A_C(\bm\epsilon)}{\partial\epsilon_j}
&=H(U\mid E_{\sim j},Y_{\sim j})
-H(U\mid E_{\sim j},Y_{\sim j},U_j)\nonumber\\
&=H(U_j\mid E_{\sim j},Y_{\sim j}).
\end{align}
On the diagonal $\epsilon_1=\cdots=\epsilon_n=\epsilon$, the total derivative yields
\begin{align*}
\frac{d}{d\epsilon}A_C(\epsilon)
=\sum_{j=1}^n H(U_j\mid E_{\sim j},Y_{\sim j})
=nh_C(\epsilon).
\end{align*}
Moreover,
\begin{align*}
A_C(\bm\epsilon)
=\sum_{\mathcal E\subseteq[n]}\dim C(\mathcal E)
\prod_{j\in\mathcal E}\epsilon_j
\prod_{j\notin\mathcal E}(1-\epsilon_j)
\end{align*}
is a polynomial in $\bm\epsilon$. The derivative identity therefore extends to the interval boundaries by one-sided derivatives.
At $\epsilon=0$, the codeword is determined, whereas at $\epsilon=1$ no coordinate is observed.
Thus, $A_C(0)=0$ and $A_C(1)=\dim C$.
The fundamental theorem of calculus gives
\begin{align*}
\int_0^1h_C(\epsilon)\,d\epsilon
=\frac{A_C(1)-A_C(0)}{n}
=\frac{\dim C}{n}.
\end{align*}
\end{proof}

\begin{proof}[Proof of Theorem~\ref{thm:gexit_bsc_area}]
Fix $j$ and hold the other crossover probabilities fixed.
The entropy chain rule gives
\begin{align*}
H(U\mid Y)
=H(U\mid Y_{\sim j})+H(Y_j\mid U,Y_{\sim j})-H(Y_j\mid Y_{\sim j}).
\end{align*}
The first term is independent of $p_j$.
Memorylessness gives $H(Y_j\mid U,Y_{\sim j})=h_2(p_j)$.
Since $Q_j=\Pr(U_j=1\mid Y_{\sim j})$, we have
\begin{align}\label{eq:extrinsic-posterior}
\Pr(Y_j=1\mid Y_{\sim j})
=p_j(1-Q_j)+(1-p_j)Q_j=p_j\cnop Q_j.
\end{align}
Thus,
\begin{align*}
G_C(\bp)
=H(U\mid Y_{\sim j})+h_2(p_j)
-\mathbb E[h_2(p_j\cnop Q_j)].
\end{align*}
The distribution of $Q_j$ does not depend on $p_j$.
Since $Y_{\sim j}$ takes finitely many values, differentiating term by term for $0<p_j<1/2$ gives
\begin{align}\label{eq:partial-derivative-bsc-gexit}
\frac{\partial G_C(\bp)}{\partial p_j}
=h_2'(p_j)-\mathbb E[(1-2Q_j)h_2'(p_j\cnop Q_j)].
\end{align}
Summing the coordinate derivatives on the diagonal proves $G_C'(p)=ng_C(p)$.

Since $h_2'(p)\to\infty$ as $p\downarrow0$, the area integral is improper at $p=0$ and we justify it directly.
For $0<\delta<1/2$, the entropy is continuously differentiable on $[\delta,1/2]$, with a one-sided derivative at $1/2$.
Consequently,
\begin{align*}
\int_\delta^{1/2}g_C(p)\,dp
=\frac{G_C(1/2)-G_C(\delta)}{n}.
\end{align*}
Increasing the crossover probability degrades the channel.
Data processing therefore makes $G_C$ nondecreasing and $g_C$ nonnegative in the interior.
Finite-alphabet entropy is continuous at $p=0$, where $G_C(0)=0$.
At $p=1/2$, $Y$ is independent of $U$ and $G_C(1/2)=\dim C$.
Taking $\delta\downarrow0$ proves the integral identity.
\end{proof}

\section{CSS GEXIT Area Theorem for Independent Bit- and Phase-Flip Noise}\label{sec:css}

The CSS notation and logical quotient are defined in Section~\ref{sec:css-setup}, and the erasure EXIT area theorem is proved in Section~\ref{sec:css-erasure}.
This appendix gives the GEXIT area theorem for independent bit- and phase-flip noise.

Suppose bit flips and phase flips occur independently.
In coordinate $j$, let the bit-error indicator $x_j$ and the phase-error indicator $z_j$ be independent Bernoulli random variables with parameter $p_j$, independently across coordinates.
As in the erasure setting, $S_Z=H_Zx$ contains the information used to determine $M_X$, while $S_X=H_Xz$ contains the information used to determine $M_Z$.
We define
\begin{align*}
    G_{\mathrm{CSS}(C_X,C_Z)}(\bp)
    \coloneqq H(M_X,M_Z\mid S_Z,S_X).
\end{align*}
When every coordinate has flip probability $p$, we write $G_{\mathrm{CSS}(C_X,C_Z)}(p)$ and define
\begin{align*}
    g_{\mathrm{CSS}(C_X,C_Z)}(p)
    \coloneqq
    \frac{1}{2n}\frac{d}{dp}
    G_{\mathrm{CSS}(C_X,C_Z)}(p).
\end{align*}

The independence assumption is essential for the decomposition below.
For the depolarizing channel with total error probability $p$, the bit- and phase-error indicators each have marginal probability $2p/3$, while their joint event is the $Y$ error and has probability $p/3$.
They are therefore correlated, and depolarizing noise must be treated through the joint four-valued Pauli-error distribution as in Theorem~\ref{thm:stabilizer-depolarizing-area}.

\begin{theorem}[CSS bit-flip and phase-flip GEXIT area theorem]\label{thm:bit-flip-gexit-area}
For a CSS code defined by \(C_X\) and \(C_Z\) with $k=\dim C_X+\dim C_Z-n$ encoded qubits, we have
\begin{align*}
    g_{\mathrm{CSS}(C_X,C_Z)}(p)
    =\frac{1}{2}\left(g_{C_Z/C_X^\perp}(p)+g_{C_X/C_Z^\perp}(p)\right)
    \quad\text{and}\quad
    \int_0^{1/2} g_{\mathrm{CSS}(C_X,C_Z)}(p)\,dp=\frac{k}{n}.
\end{align*}
In the normalized noise-entropy parameter $a=h_2(p)$, the function
\begin{align*}
    \gamma_{\mathrm{CSS}(C_X,C_Z)}(a)
    \coloneqq\frac{1}{2n}\frac{d}{da}G_{\mathrm{CSS}(C_X,C_Z)}\bigl(h_2^{-1}(a)\bigr)
    =\frac{1}{2}\left(\gamma^{\mathrm{BSC}}_{C_Z/C_X^\perp}(a)+\gamma^{\mathrm{BSC}}_{C_X/C_Z^\perp}(a)\right)
\end{align*}
satisfies $0\leq\gamma_{\mathrm{CSS}(C_X,C_Z)}(a)\leq1$ for $0<a<1$ and $\int_0^1\gamma_{\mathrm{CSS}(C_X,C_Z)}(a)\,da=k/n$.
\end{theorem}

\begin{proof}
Under independent bit- and phase-flip noise, $x$ and $z$ are independent binary error vectors.
Their syndromes $S_Z=H_Zx$ and $S_X=H_Xz$ depend on separate components.
Choose the $x$ correction as a function of $S_Z$ and the $z$ correction as a function of $S_X$.
The resulting logical classes $M_X$ and $M_Z$ are independent conditioned on the two syndromes.
Consequently,
\begin{align*}
H(M_X,M_Z\mid S_Z,S_X)
=H(M_X\mid S_Z)+H(M_Z\mid S_X).
\end{align*}
Applying the argument of Lemma~\ref{lem:syndrome} to the binary nested pairs $C_X^\perp\subseteq C_Z$ and $C_Z^\perp\subseteq C_X$ shows that each syndrome posterior is a translate of the corresponding coset posterior on a BSC.
Hence, $G_{\mathrm{CSS}(C_X,C_Z)}(\bp)=G_{C_Z/C_X^\perp}(\bp)+G_{C_X/C_Z^\perp}(\bp)$.

Differentiating the diagonal decomposition and dividing by $2n$ yields
\begin{align*}
g_{\mathrm{CSS}(C_X,C_Z)}(p)
=\frac12\left(g_{C_Z/C_X^\perp}(p)+g_{C_X/C_Z^\perp}(p)\right),
\qquad 0<p<1/2.
\end{align*}
Both component functions are integrable by Theorem~\ref{thm:coset_gexit_bsc_area}.
Each component quotient has dimension $k$, so integration gives
\begin{align*}
\int_0^{1/2}g_{\mathrm{CSS}(C_X,C_Z)}(p)\,dp
=\frac12\left(\frac{k}{n}+\frac{k}{n}\right)=\frac{k}{n}.
\end{align*}
The statements for $\gamma_{\mathrm{CSS}(C_X,C_Z)}$ follow from \eqref{eq:bsc-entropy-gexit}, Lemma~\ref{lem:final1-bsc-coset-pointwise}, and the change of variables $a=h_2(p)$.
\end{proof}

\section{Details of the Numerical Evaluation}\label{app:numerical}

This appendix gives additional details for the numerical computations in Sections~\ref{sec:numerical-mc} and~\ref{sec:numerical-pauli}.

\subsection{Rank Profiles of Nested Erasure Sets}\label{app:numerical-nested}
The parity-check matrices $H^{C_Z}$ and $H^{C_X}$ in \eqref{eq:bb-mc-class-x} are obtained from $H_Z$ and $H_X$ by removing linearly dependent rows, which does not change the rank of any set of columns, so all ranks are computed directly from the columns of $H_Z$ and $H_X$.
For each fixed $\epsilon$, every coordinate belongs to $E(\epsilon)$ independently with probability $\epsilon$, which is the erasure pattern of the quantum erasure channel in Section~\ref{sec:pauli-channels}.
Inserting the columns $\pi(1),\ldots,\pi(n)$ one at a time and updating the rank after each insertion gives $\rank H^{C_Z}_{E_t}$ and $\rank H^{C_X}_{E_t}$ for all $t=0,\ldots,n$.
Inserting them in the order $\pi(n),\ldots,\pi(1)$ gives the ranks for the complements $\overline{E_t}$.
With these four rank profiles, one sample gives the exact correction-class entropies of both components for all $n+1$ sets $E_t$.
For the bivariate-bicycle codes, the rank updates use Gaussian elimination over $\bbF_2$ that inserts one column per step.
For the planar surface codes, they use union-find as described in Appendix~\ref{app:numerical-surface}.

\subsection{Estimators and Standard Errors}\label{app:numerical-estimators}
Since $\pi$ is a uniformly random permutation, $E_t$ is uniformly distributed over the subsets of $[n]$ of size $t$, which makes $\hat\alpha_t$ an unbiased estimate of $\alpha_t$.
The identity \eqref{eq:bb-mc-bernstein-derivative} also agrees with the average over $j$ of \eqref{eq:final1-bec-coset-slope}, because $\alpha_{t+1}-\alpha_t$ is the expected increase of the correction-class entropy when a uniformly chosen coordinate is added to a uniformly random erasure set of size $t$.
We evaluate \eqref{eq:bb-mc-bernstein} and \eqref{eq:bb-mc-bernstein-derivative} at $\epsilon\in\{0,0.001,\ldots,1\}$.

The samples of each code are divided into independent batches of equal size, which are listed in Tables~\ref{tab:surface-mc} and~\ref{tab:bb-mc}.
Each batch uses an independent random draw.
The standard error of each reported value is $s/\sqrt{N_{\mathrm b}}$, where $N_{\mathrm b}$ is the number of batches and $s$ is the sample standard deviation of the $N_{\mathrm b}$ batch estimates.
Over $\epsilon\in[0,1]$, the largest standard error of $A_{C_Z/C_X^\perp}(\epsilon)/k$ is $2.2\times10^{-4}$ for the surface codes and $7.6\times10^{-5}$ for the bivariate-bicycle codes, and the largest standard error of $h_{C_Z/C_X^\perp}(\epsilon)$ is $1.8\times10^{-6}$ and $1.9\times10^{-5}$, respectively.
Each sample also records \eqref{eq:bb-mc-class-x} at $\cE=E(\epsilon)$ for $\epsilon\in\{0,0.0025,\ldots,1\}$.
The sample means of these values estimate $A_{C_Z/C_X^\perp}(\epsilon)$ without \eqref{eq:bb-mc-bernstein}.
Over the $401$ values of $\epsilon$, they differ from \eqref{eq:bb-mc-bernstein} by at most $2.48$ standard errors for every surface code and by at most $2.82$, $3.74$, $2.17$, and $2.89$ standard errors for $[[576,16]]$, $[[2304,16]]$, $[[1152,12]]$, and $[[4608,12]]$.

\subsection{Planar Surface Codes}\label{app:numerical-surface}
Let $H_{\mathrm{rep}}\in\bbF_2^{(d-1)\times d}$ be the parity-check matrix of the length-$d$ repetition code whose $i$th row has ones in positions $i$ and $i+1$.
The planar surface code of distance $d$ has the parity-check matrices
\begin{align*}
    H_X=[H_{\mathrm{rep}}\otimes I_d\mid I_{d-1}\otimes H_{\mathrm{rep}}^T],
    \qquad
    H_Z=[I_d\otimes H_{\mathrm{rep}}\mid H_{\mathrm{rep}}^T\otimes I_{d-1}],
\end{align*}
so that $n=d^2+(d-1)^2$, $\rank H_X=\rank H_Z=d(d-1)$, and $k=1$.
The first $d^2$ coordinates are indexed by pairs in $[d]\times[d]$ and the last $(d-1)^2$ coordinates by pairs in $[d-1]\times[d-1]$, following the Kronecker products.
The permutation $\pi_{XZ}$ of $[n]$ that maps the pair $(i,j)$ to $(j,i)$ within each of the two blocks is an involution, and it maps the row space of $H_X$ onto the row space of $H_Z$, which we verified for $d\in\{3,5,7,11\}$.
It follows that $\pi_{XZ}$ maps $C_X$ to $C_Z$ and $C_X^\perp$ to $C_Z^\perp$.

\begin{figure}[t]
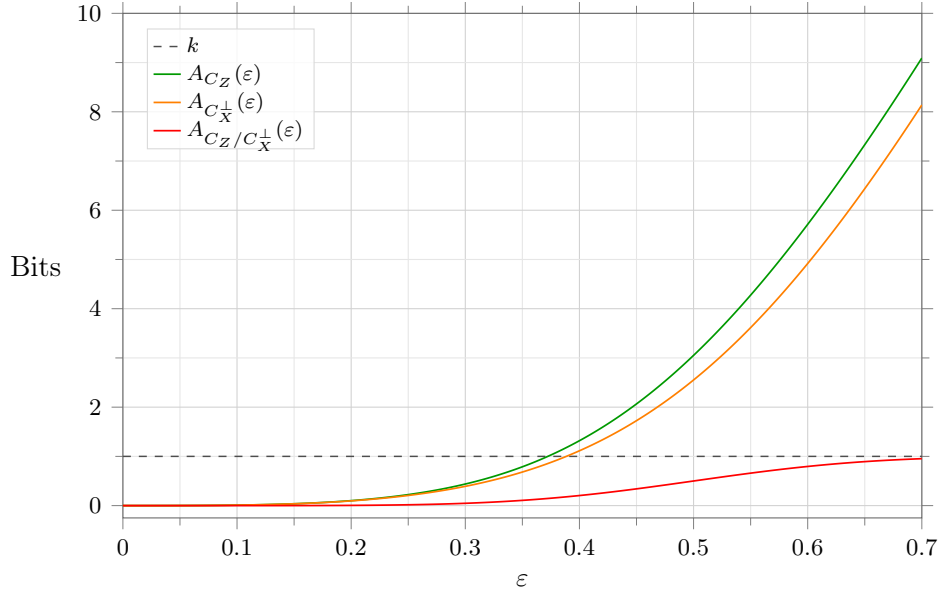

\centering

 \caption{Correction-class decomposition $A_{C_Z}(\epsilon)=A_{C_Z/C_X^{\perp}}(\epsilon)+A_{C_X^{\perp}}(\epsilon)$ for the $[[41,1,5]]$ planar surface code on the quantum erasure channel, estimated as described in Section~\ref{sec:numerical-mc}. The dashed line marks $k=1$, the number of encoded logical qubits.}
\label{fig:mc-decomposition}
\end{figure}
 
Fig.~\ref{fig:mc-decomposition} shows that, for the $[[41,1,5]]$ code at $\epsilon=1/2$, the ambiguity $A_{C_X^\perp}(1/2)=2.552$ among stabilizer-equivalent errors accounts for most of $A_{C_Z}(1/2)=3.052$, while the correction-class entropy is $A_{C_Z/C_X^\perp}(1/2)=0.500$.

Every column of $H_Z$ has weight one or two.
Let $\cG_Z$ be the graph whose vertices are the rows of $H_Z$ together with one boundary vertex, in which coordinate $j$ is an edge that joins the two rows in the support of column $j$, or the single row in its support and the boundary vertex.
Each column of $H_Z$ is the incidence vector of its edge with the entry of the boundary vertex removed, so $\rank H^{C_Z}_{\cE}$ equals the number of edges in a spanning forest of the edges in $\cE$.
Union-find inserts the edges $\pi(1),\ldots,\pi(n)$ and increases the rank exactly when the two endpoints of the inserted edge lie in different components.
The same construction applies to $H_X$.

For $k=1$, \eqref{eq:bb-mc-class-x} takes the values $0$ and $1$.
It equals $1$ exactly when the erased edges contain a path in $\cG_Z$ that joins the two opposite sides of the lattice at which the boundary edges are attached, and such a path is the support of a nontrivial logical $X$ operator.
$A_{C_Z/C_X^\perp}(\epsilon)$ is the probability of such a crossing in bond percolation on this lattice with bond probability $\epsilon$~\cite{Stace-prl09}, and the critical bond probability of the infinite square lattice is $1/2$~\cite{Kesten-cmp80}.
The symmetry $A_{C_Z/C_X^\perp}(\epsilon)+A_{C_Z/C_X^\perp}(1-\epsilon)=k$ gives $A_{C_Z/C_X^\perp}(1/2)=1/2$ for every $d$, and the estimates in Table~\ref{tab:surface-mc} agree with this value within $2.52$ standard errors.
A least-squares fit of the logarithm of the width of the interval in Table~\ref{tab:surface-mc} against $\log d$ gives the slope $-0.745$ over $d\ge21$ and $-0.746$ over $d\ge45$.
These slopes agree with the finite-size scaling $d^{-1/\nu}$ with the correlation-length exponent $\nu=4/3$ of two-dimensional percolation~\cite{Stauffer-1994}.

\begin{table}[t]
\centering
\caption{Monte Carlo results for the $X$ component of the planar surface codes on the quantum erasure channel. The interval lists the erasure probabilities at which $A_{C_Z/C_X^\perp}(\epsilon)$ equals $0.1$ and $0.9$, and the digits in parentheses are the standard error in the last digits.}
\label{tab:surface-mc}
\footnotesize
\begin{tabular}{ccccccc}
\hline
$d$ & $n$ & Batches $\times$ samples & $h_{C_Z/C_X^\perp}(1/2)$ & $1/h_{C_Z/C_X^\perp}(1/2)$ & $A_{C_Z/C_X^\perp}(1/2)$ & Interval\\
\hline
$5$ & $41$ & $20\times5\times10^{7}$ & $0.0799734(12)$ & $12.50$ & $0.50000(1)$ & $[0.3458,0.6542]$\\
$7$ & $85$ & $20\times5\times10^{7}$ & $0.0497477(11)$ & $20.10$ & $0.50001(1)$ & $[0.3793,0.6208]$\\
$11$ & $221$ & $20\times3\times10^{7}$ & $0.0268447(7)$ & $37.25$ & $0.50001(2)$ & $[0.4133,0.5866]$\\
$21$ & $841$ & $30\times10^{7}$ & $0.0114228(5)$ & $87.54$ & $0.50002(2)$ & $[0.4463,0.5537]$\\
$45$ & $3961$ & $30\times2.5\times10^{6}$ & $0.0042805(5)$ & $233.6$ & $0.49995(6)$ & $[0.4695,0.5305]$\\
$75$ & $11101$ & $40\times8\times10^{5}$ & $0.0022369(4)$ & $447.0$ & $0.50017(7)$ & $[0.4792,0.5208]$\\
$101$ & $20201$ & $40\times4\times10^{5}$ & $0.0015361(5)$ & $651.0$ & $0.49988(9)$ & $[0.4833,0.5167]$\\
$151$ & $45301$ & $60\times1.5\times10^{5}$ & $0.0009250(4)$ & $1081$ & $0.50017(13)$ & $[0.4876,0.5124]$\\
$251$ & $125501$ & $120\times3.75\times10^{4}$ & $0.0004886(4)$ & $2046$ & $0.49992(22)$ & $[0.4915,0.5085]$\\
\hline
\end{tabular}
\end{table}

\subsection{Bivariate Bicycle Codes}\label{app:numerical-bb}
The bivariate bicycle codes are constructed as in~\cite{Bravyi-nature24}.
For positive integers $\ell_1$ and $\ell_2$, let $\mathsf{S}_{\ell}$ denote the $\ell\times\ell$ cyclic shift matrix, and let $\mathsf{x}=\mathsf{S}_{\ell_1}\otimes I_{\ell_2}$ and $\mathsf{y}=I_{\ell_1}\otimes\mathsf{S}_{\ell_2}$ be the corresponding commuting $\ell_1\ell_2\times\ell_1\ell_2$ permutation matrices.
For two polynomials $\mathsf{A}$ and $\mathsf{B}$ in $\mathsf{x}$ and $\mathsf{y}$, the parity-check matrices are
\begin{align*}
    H_X=[\mathsf{A}\mid\mathsf{B}],
    \qquad
    H_Z=[\mathsf{B}^T\mid\mathsf{A}^T],
\end{align*}
so that $n=2\ell_1\ell_2$ and $H_XH_Z^T=\mathsf{A}\mathsf{B}+\mathsf{B}\mathsf{A}=\mzero$ over $\bbF_2$.
The codes $[[576,16]]$ and $[[2304,16]]$ use $\mathsf{A}=\mathsf{x}^3+\mathsf{y}+\mathsf{y}^2$ and $\mathsf{B}=\mathsf{y}^3+\mathsf{x}+\mathsf{x}^2$ with $(\ell_1,\ell_2)=(24,12)$ and $(\ell_1,\ell_2)=(48,24)$.
The codes $[[1152,12]]$ and $[[4608,12]]$ use $\mathsf{A}=\mathsf{x}^3+\mathsf{y}^2+\mathsf{y}^7$ and $\mathsf{B}=\mathsf{y}^3+\mathsf{x}+\mathsf{x}^2$ with $(\ell_1,\ell_2)=(24,24)$ and $(\ell_1,\ell_2)=(48,48)$.
For $(\ell_1,\ell_2)=(12,6)$ and $(\ell_1,\ell_2)=(12,12)$, the same two pairs of polynomials give the $[[144,12,12]]$ and $[[288,12,18]]$ codes of~\cite{Bravyi-nature24}.
For each code, $\rank H_X=\rank H_Z$ takes the value listed in Table~\ref{tab:bb-mc}, and the number of encoded qubits is $k=n-\rank H_X-\rank H_Z=\dim C_X+\dim C_Z-n$.

Let $\pi_0$ be the permutation of the $\ell_1\ell_2$ indices that maps $(i,j)\in\mathbb Z_{\ell_1}\times\mathbb Z_{\ell_2}$ to $(-i,-j)$, and let $\pi_{XZ}$ be the permutation of $[n]$ that exchanges the two halves of the coordinates and applies $\pi_0$ within each half.
The permutation matrix of $\pi_0$ conjugates each monomial $\mathsf{x}^{i}\mathsf{y}^{j}$ to its inverse, which equals its transpose.
It follows that $\pi_{XZ}$ maps the row space of $H_X$ onto the row space of $H_Z$, and since $\pi_{XZ}$ is an involution, it maps $C_X^\perp$ to $C_Z^\perp$, $C_Z^\perp$ to $C_X^\perp$, and $C_X$ to $C_Z$.
We verified the equality of the two row spaces for $[[576,16]]$ and $[[1152,12]]$.

\begin{table}[t]
\centering
\caption{Monte Carlo results for the $X$ component of the bivariate-bicycle codes on the quantum erasure channel. The interval lists the erasure probabilities at which $A_{C_Z/C_X^\perp}(\epsilon)/k$ equals $0.1$ and $0.9$, and the digit in parentheses is the standard error in the last digit.}
\label{tab:bb-mc}
\footnotesize
\begin{tabular}{cccccccc}
\hline
Code & $(\ell_1,\ell_2)$ & $\rank H_X$ & Batches $\times$ samples & $h_{C_Z/C_X^\perp}(1/2)$ & $1/h_{C_Z/C_X^\perp}(1/2)$ & Interval & Distance\\
\hline
$[[576,16]]$ & $(24,12)$ & $280$ & $60\times8\times10^{5}$ & $0.428604(5)$ & $2.333$ & $[0.4669,0.5331]$ & $\le20$\\
$[[2304,16]]$ & $(48,24)$ & $1144$ & $80\times6\times10^{4}$ & $0.238197(8)$ & $4.198$ & $[0.4851,0.5149]$ & $\le40$\\
$[[1152,12]]$ & $(24,24)$ & $570$ & $60\times2.5\times10^{5}$ & $0.254994(4)$ & $3.922$ & $[0.4791,0.5209]$ & $\le38$\\
$[[4608,12]]$ & $(48,48)$ & $2298$ & $200\times10^{4}$ & $0.132509(6)$ & $7.547$ & $[0.4900,0.5101]$ & $\le182$\\
\hline
\end{tabular}
\end{table}

In each sample, let $t^\ast$ be the smallest $t$ for which \eqref{eq:bb-mc-class-x} is positive at $\cE=E_t$.
Since $C_X^\perp(\cE)\subseteq C_Z(\cE)$ with equality at $\cE=E_{t^\ast-1}$, adding the coordinate $\pi(t^\ast)$ increases $\dim C_Z(\cE)$ by one and leaves $\dim C_X^\perp(\cE)$ unchanged.
The column of $H_Z$ indexed by $\pi(t^\ast)$ is a sum of columns indexed by a subset of $E_{t^\ast-1}$, and the corresponding vector $\bv\in C_Z(E_{t^\ast})$ satisfies $\bv\notin C_X^\perp$.
It represents a nontrivial logical $X$ operator.
We reduce its weight by adding rows of $H_X$ as long as the weight decreases, which keeps $\bv$ in the coset $\bv+C_X^\perp$.
The smallest weight found over all samples is an upper bound on the minimum weight of the vectors in $C_Z\setminus C_X^\perp$, and the same procedure with $X$ and $Z$ exchanged bounds the minimum weight of the vectors in $C_X\setminus C_Z^\perp$.
The bounds are $20$ and $20$ for $[[576,16]]$, $42$ and $40$ for $[[2304,16]]$, $38$ and $38$ for $[[1152,12]]$, and $182$ and $190$ for $[[4608,12]]$, which gives the distance bounds in Table~\ref{tab:bb-mc}.
With $300$ samples, the procedure finds weights $12$ and $12$ for the $[[144,12,12]]$ code and $22$ and $18$ for the $[[288,12,18]]$ code.
For all six codes, each vector that attains a reported bound was verified to lie in $C_Z\setminus C_X^\perp$ or in $C_X\setminus C_Z^\perp$, respectively.
For $[[4608,12]]$, the bound is far from tight, because $t^\ast$ was at least $2210$ in every sample and the greedy reduction removes only local redundancy.

\vspace{-1.5mm}
\subsection{Small Codes}\label{app:numerical-small}

\begin{figure}[t]
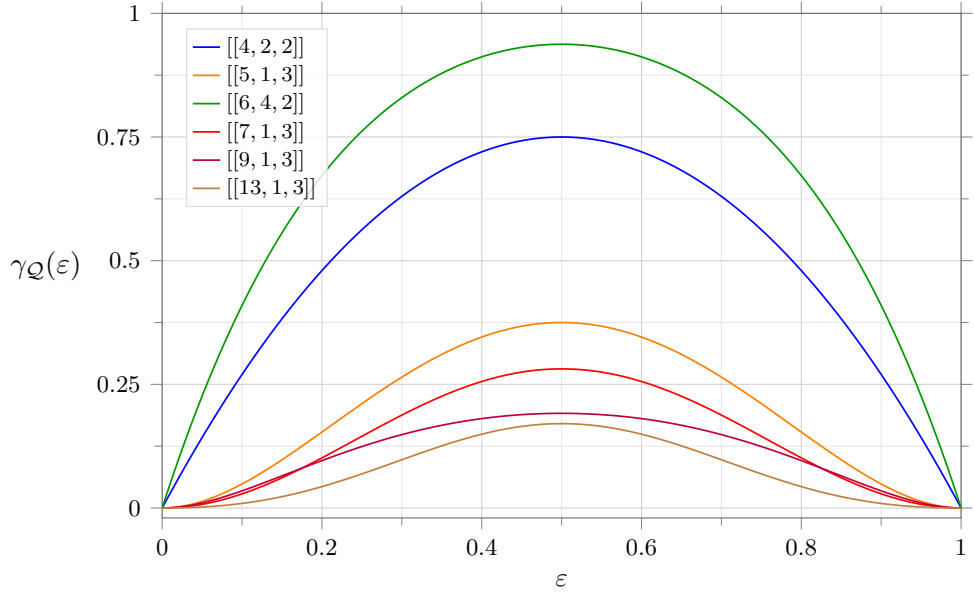

\centering

 \caption{Stabilizer EXIT function $\gamma_{\cQ}(\epsilon)$ on the quantum erasure channel for the small codes in Fig.~\ref{fig:pauli-exact}, computed as described in Appendix~\ref{app:numerical-small}. Each area equals $k/n$, and every curve lies in $[0,1]$.}
\label{fig:small-erasure}
\end{figure}
 \begin{figure}[t]
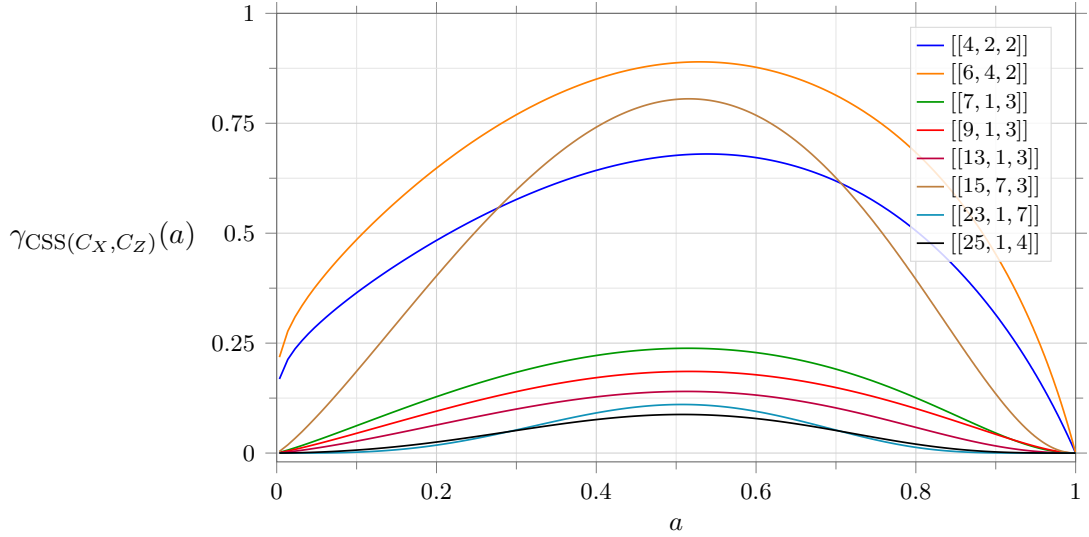

\centering
%
 \caption{CSS GEXIT function $\gamma_{\mathrm{CSS}(C_X,C_Z)}(a)$ of small CSS codes under independent bit- and phase-flip noise with $a=h_2(p)$, computed as described in Appendix~\ref{app:numerical-small}. Each area equals $k/n$, and every curve lies in $[0,1]$.}
\label{fig:small-bitflip}
\end{figure}
 
The $[[4,2,2]]$ and $[[6,4,2]]$ codes are the CSS codes with $H_X=H_Z=[1\,1\,1\,1]$ and $H_X=H_Z=[1\,1\,1\,1\,1\,1]$.
The $[[5,1,3]]$ code is the stabilizer code generated by $XZZXI$ and its cyclic shifts $IXZZX$, $XIXZZ$, and $ZXIXZ$, and it is not a CSS code.
The Steane $[[7,1,3]]$ code is the CSS code with $H_X=H_Z$ equal to the parity-check matrix of the $[7,4]$ Hamming code.
The Shor $[[9,1,3]]$ code is the CSS code whose $Z$-type stabilizers are $Z_jZ_{j+1}$ for $j\in\{1,2,4,5,7,8\}$ and whose $X$-type stabilizers are $X_1\cdots X_6$ and $X_4\cdots X_9$.
Its stabilizers of weight $2$ are lighter than its distance $3$, so it is degenerate.
The $[[13,1,3]]$ and $[[25,1,4]]$ codes are the planar surface codes of Appendix~\ref{app:numerical-surface} with $d=3$ and $d=4$.
The $[[15,7,3]]$ code has $H_X=H_Z$ equal to the parity-check matrix of the $[15,11]$ Hamming code.
The $[[23,1,7]]$ code has $H_X=H_Z$ equal to a generator matrix of the $[23,11]$ dual of the cyclic Golay code with generator polynomial $1+D^2+D^4+D^5+D^6+D^{10}+D^{11}$, and this dual is contained in the Golay code.

\subsubsection{Depolarizing channel}
Each Pauli-error vector $\be\in(\bbF_2^2)^n$ is mapped to its syndrome, given by its symplectic inner products with the rows of a generator matrix of $\cS$, and to a class label, given by its symplectic inner products with $2k$ vectors that extend a basis of $\cS$ to a basis of $\cS^\perp$.
Two Pauli-error vectors with the same syndrome differ by an element of $\cS^\perp$, and this element lies in $\cS=(\cS^\perp)^\perp$ exactly when its symplectic inner products with these $2k$ vectors vanish.
The class label identifies the logical correction class $M$ of Section~\ref{sec:syndrome-coset} up to a fixed relabeling, which does not change $H(M\mid S)$.
The depolarizing channel applies $\be$ with probability $(1-p)^{n-i}(p/3)^{i}$, where $i$ is the number of coordinates $j$ with $\be_j\neq0$.
Let $N_{s,m,i}$ be the number of Pauli-error vectors with syndrome $s$, class label $m$, and $i$ nonzero coordinates, which one enumeration of all $4^n$ vectors gives.
Then
\begin{align*}
    \Pr(S=s,M=m)&=\sum_{i=0}^{n}N_{s,m,i}(1-p)^{n-i}\left(\frac{p}{3}\right)^{i},\\
    G_{\cQ}(a(p))&=-\sum_{s,m}\Pr(S=s,M=m)\log_2\frac{\Pr(S=s,M=m)}{\Pr(S=s)},\\
    \frac{d}{dp}G_{\cQ}(a(p))&=-\sum_{s,m}\frac{d\Pr(S=s,M=m)}{dp}\log_2\frac{\Pr(S=s,M=m)}{\Pr(S=s)}.
\end{align*}
The last identity holds because the terms obtained by differentiating the logarithm add up to a multiple of $\frac{d}{dp}\sum_{s,m}\Pr(S=s,M=m)-\frac{d}{dp}\sum_{s}\Pr(S=s)=0$.
With $\frac{d}{dp}a(p)=\frac12\log_2\frac{3(1-p)}{p}$, the stabilizer GEXIT function is $g_{\cQ}(a(p))=\frac{1}{2n}\frac{d}{dp}G_{\cQ}(a(p))\big/\frac{d}{dp}a(p)$.
We evaluate these expressions at $1501$ equally spaced values of $p$ in $[10^{-6},3/4]$.
For the $[[13,1,3]]$ code, the enumeration covers $4^{13}\approx6.7\times10^7$ Pauli-error vectors.
These expressions give the entropy of the posterior distribution of logical correction classes conditioned on the syndrome, along with its derivative. The MAP decoder selects the most probable class under this distribution.

For the five codes with $n\le9$, the code space is the joint $+1$ eigenspace of the Hermitian Pauli operators that represent the rows of a generator matrix of $\cS$, and the state $\ket{\Omega}_{L'B^nF^n}$ of Section~\ref{sec:converse} is formed with an orthonormal basis of this space.
Applying $\cN_p$ to each of the $n$ qubits gives the state of $L'B^n$, whose dimension is at most $2^{k+n}=1024$, and $I_c(L'\rangle B^n)=H(B^n)-H(L'B^n)$ follows from the eigenvalues of the density matrices of $L'B^n$ and $B^n$.
We compared $k-I_c(L'\rangle B^n)$ with $G_{\cQ}(a(p))$ at $p\in\{0.02,0.1,0.1893,0.25,0.4,0.6,0.74\}$.

\vspace{-1.5mm}
\subsubsection{Independent bit- and phase-flip noise}
For the $X$ component of a CSS code under independent bit- and phase-flip noise, each bit-error vector $x\in\bbF_2^n$ is mapped to its syndrome $H_Zx$ and to a class label, given by its inner products with $k$ vectors that extend a basis of $C_Z^\perp$ to a basis of $C_X$.
Two bit-error vectors with the same syndrome differ by an element $c\in C_Z$, and $c$ lies in $C_X^\perp$ exactly when it is orthogonal to these $k$ vectors, because $c$ is orthogonal to $C_Z^\perp$.
The bit-error vector $x$ has probability $(1-p)^{n-i}p^{i}$, where $i$ is the number of ones in $x$, and the same three expressions as above with $p/3$ replaced by $p$ give $H(M_X\mid S_Z)$ and its derivative.
The $Z$ component follows by exchanging the roles of $X$ and $Z$, and Theorem~\ref{thm:bit-flip-gexit-area} gives $G_{\mathrm{CSS}(C_X,C_Z)}(p)=H(M_X\mid S_Z)+H(M_Z\mid S_X)$.
We evaluate $\gamma_{\mathrm{CSS}(C_X,C_Z)}(a)=\frac{1}{2n}\frac{d}{dp}G_{\mathrm{CSS}(C_X,C_Z)}(p)\big/\log_2\frac{1-p}{p}$ with $a=h_2(p)$ at $2001$ equally spaced values of $p$ in $[10^{-6},1/2]$, and we evaluate each component $\gamma^{\mathrm{BSC}}_{C_Z/C_X^\perp}(a)$ and $\gamma^{\mathrm{BSC}}_{C_X/C_Z^\perp}(a)$ in the same way.
Fig.~\ref{fig:small-bitflip} shows $\gamma_{\mathrm{CSS}(C_X,C_Z)}(a)$ for the eight CSS codes.
For all eight codes, $G_{\mathrm{CSS}(C_X,C_Z)}(1/2)=2k$ up to $2\times10^{-15}$, and $\gamma_{\mathrm{CSS}(C_X,C_Z)}(a)$ and its two components lie in $[0,1]$ up to rounding errors below $10^{-11}$, as Theorem~\ref{thm:bit-flip-gexit-area} and Lemma~\ref{lem:final1-bsc-coset-pointwise} state.
The largest values $0.890$ and $0.806$ occur for the $[[6,4,2]]$ and $[[15,7,3]]$ codes, whose rates are $2/3$ and $7/15$.
For the Shor code, whose $X$- and $Z$-type stabilizers differ, the components differ, and the largest value of $\gamma^{\mathrm{BSC}}_{C_Z/C_X^\perp}$ is $0.192$.

\vspace{-1.5mm}
\subsubsection{Quantum erasure channel}
For a fixed erasure set $\cE$, the correction-class entropy is $H(M\mid S,E=\cE)=\dim\cS^\perp(\cE)-\dim\cS(\cE)$, as shown before Theorem~\ref{thm:folding}.
For an additive code $C\subseteq(\bbF_2^2)^n$, $\dim C(\cE)$ equals $\dim C$ minus the rank of the restriction of $C$ to the $2|\overline{\cE}|$ coordinates of $\overline{\cE}$.
For the six codes of Fig.~\ref{fig:pauli-exact}, we enumerate all $2^n$ erasure sets and average this entropy over the erasure sets of each size.
The functions $A_{\cQ}(\epsilon)$ and $\gamma_{\cQ}(\epsilon)$ then follow from these averages as in \eqref{eq:bb-mc-bernstein} and \eqref{eq:bb-mc-bernstein-derivative}, with the normalization $1/(2n)$ of $\gamma_{\cQ}$.
Fig.~\ref{fig:small-erasure} shows $\gamma_{\cQ}(\epsilon)$.
For all six codes, $A_{\cQ}(0)=0$ and $A_{\cQ}(1)=2k$, and $\gamma_{\cQ}(\epsilon)$ lies in $[0,1]$, as Theorem~\ref{thm:stabilizer-erasure-area} states.
The identity $H(M\mid S,E=\cE)+H(M\mid S,E=\overline{\cE})=2k$ from the proof of Theorem~\ref{thm:folding} holds for every erasure set, and $\gamma_{\cQ}(\epsilon)$ is symmetric about $\epsilon=1/2$ up to rounding errors below $10^{-15}$.
The maximum of $\gamma_{\cQ}$ occurs at $\epsilon=1/2$ for every code.

\end{document}